\RequirePackage{fix-cm}
\documentclass[natbib]{svjour3}                     
\smartqed  
\usepackage{colortbl}
\usepackage[utf8]{inputenc}
\usepackage[T1]{fontenc} 
\usepackage{graphicx}
\usepackage{url}
\usepackage{arydshln}
\usepackage{todonotes}
\usepackage{rotating}

\newcommand{\RQOne}{Does everyone in the development team share the same level of well-being?}
\newcommand{\RQTwo}{Can software developers' actions predict well-being?}
\newcommand{\RQThree}{Can software developers' well-being and actions predict software developers’ productivity?}
\newcommand{\RQFour}{Can interviews give further information about experienced well-being of software developers?}
\newcommand{\eg}{\emph{e.g.}}
\newcommand{\ie}{\emph{i.e.}}

\begin{document}



\title{Individual Differences Limit Predicting Well-being and Productivity Using Software Repositories: A Longitudinal Industrial Study}




\author{Miikka Kuutila  \and
        Mika M{\"a}ntyl{\"a} \and 
        Ma{\"e}lick Claes \and 
        Marko Elovainio \and 
        Bram Adams
}


\institute{M. Kuutila \at
              first address \\
              Tel.: +123-45-678910\\
              Fax: +123-45-678910\\
              \email{miikka.kuuttila@oulu.fi}
           \and
           S. Author \at
              second address
}

\date{Received: date / Accepted: date}

\maketitle

\begin{abstract}
Reports of poor work well-being and fluctuating productivity in software engineering have been reported in both academic and popular sources. Understanding and predicting these issues through repository analysis might help manage software developers' well-being.
Our objective is to link data from software repositories, that is commit activity, communication, expressed sentiments, and job events, with measures of well-being obtained with a daily experience sampling questionnaire. To achieve our objective, we studied a single software project team for eight months in the software industry. Additionally, we performed semi-structured interviews to explain our results. The acquired quantitative data are analyzed with generalized linear mixed-effects models with autocorrelation structure. We find that individual variance accounts for most of the $R^2$ values in models predicting developers' experienced well-being and productivity. In other words, using software repository variables to predict developers' well-being or productivity is challenging due to individual differences. Prediction models developed for each developer individually work better, with fixed effects $R^2$ value of up to 0.24. The semi-structured interviews give insights into the well-being of software developers and the benefits of chat interaction. Our study suggests that individualized prediction models are needed for well-being and productivity prediction in software development. 

\end{abstract}

\section{Introduction}\label{sec:introduction}

In recent years, the psychological well-being of software developers has drawn increased scientific interest from the fields of behavioral software engineering \citep{lenberg2015behavioral}, which borrows its name from the field of behavioral economics, and of ``psychoempirical software engineering'' \citep{graziotin2015understanding}. Software engineering researchers have established focused venues to study the affective states of software developers such as \textit{The International Workshop on Emotion Awareness in Software Engineering}\footnote{https://semotion.github.io/2021/}. 

%
Subjective well-being has been described as a broad range of phenomena, including people's emotional responses, domain satisfactions, and global judgments of satisfaction \citep{diener1999subjective}. People have been shown to use momentary affective states as information in judging their well-being \citep{schwarz1983mood}. Core affect has been defined as ``A neurophysiological state that is consciously accessible as a simple, nonreflective feeling that is an integral blend of hedonic (pleasure–displeasure) and arousal (sleepy–activated) values'', and ``the simplest raw (nonreflective) feelings evident in moods and emotions'' \citep{russell2003core}.
Studies on mining software repositories have made several recent attempts to build tools and to reason about the affective states of software developers by utilizing sentiment analysis (\eg, \cite{mantyla2017bootstrapping} and \cite{novielli2018benchmark}). However, to the best of our knowledge, no prior studies have attempted to link daily experience sampling of affective states with measures from software repositories in a longitudinal industrial setting. According to \cite{scollon2009experience}, strong points in experience sampling are its ability to document real-life experiences that improve ecological validity, to reduce of memory bias, and to augment of other research methods. We give more details of experience sampling in Section~\ref{sec:ESM}.

This paper investigates whether different software development actions are associated with different affective states and self-reported well-being. 
To achieve our goal, we used experience sampling methodology and created a questionnaire to be completed daily in an industrial software project setting. The questionnaire is based on psychosocial theories of work \citep{karasek1990healthy}. The questionnaire assesses hurry, stress, sleeping problems, interruptions, ineffective software development (defined as poorly working tools, processes or communication), and job control (independence). Metrics obtained with the questionnaire were then linked to measures obtained from software repositories related to code commit activity, amount of social interaction in an instant messaging application, the sentiment expressed through words, emoticons and emojis, and job events. We built generalized linear mixed effects models to understand the relations between software repository variables, which reflect software development actions, and the answers to the questionnaire. Additionally, we conducted semi-structured interviews to better understand the project context and reasons for discovering different relationships in the models. 

Hence, our research questions were formulated as:
\begin{description}
\item[RQ1] {\RQOne}
\item[RQ2] {\RQTwo}*
\item[RQ3] {\RQThree}*
\item[RQ4] {\RQFour}+ 
\end{description}

In our research questions, experienced well-being refers to our questionnaire. Our questionnaire asks developers for stress, hurry, sleeping problems, interruptions, independence, and ineffective software development. Similarly, in our research questions, software developers' actions refer to the multitude of variables mined from software repositories, \eg, commit related activity, amount of communication, expressed sentiment in communication, and job events.

This paper is an extension of our prior conference paper \citep{kuutila2018using} that analyzed RQ2 and workshop paper that investigated RQ3 \citep{kuutila2020chat} with different methodologies. \cite{kuutila2018using} looked at a limited set of count variables related to productivity and chat messages and their relationship to the questionnaire variables with logistic regression, where we also used binning for these count variables. Here we extend with more variables like sentiment analysis, customer meeting, and build failure information. Additionally, we changed the statistical analysis to a generalized linear mixed-effects model with an autocorrelation structure. This allows us to control the effect at the individual level while also accounting for the autocorrelation. 
\cite{kuutila2020chat} examines the sentiment analysis related variables in relation to lines of code and commits produced by the developers. Here we add the questionnaire responses, customer meetings and build failure information. Additionally, we use generalized linear mixed-effects with autocorrelation structure to control the effect of the individual. 

For clarity, we have marked the research questions with added variables and new statistical analysis using ``*'' in the above description. Semi-structured interviews were completely new for this extension and marked with a "+" for this reason.

The rest of the paper is structured as follows. The relevant background from psychology and software engineering is introduced in  Section~\ref{sec:background}. The methodology for creating the daily questionnaire and executing this study is explained in Section~\ref{sec:methodology}. In Section~\ref{sec:results} we present the results to our research questions and and discuss them in Section~\ref{sec:discussion}. We discuss internal and external threats in Section~\ref{sec:threats}. Lastly, conclusions are provided in Section~\ref{sec:conclusions}.


\section{Background}\label{sec:background}

\subsection{Work Well-being in Psychology}

Subjective well-being has been described as a broad category of phenomena, including people's emotional responses, domain satisfactions, and global judgments of satisfaction \citep{diener1999subjective}. Moreover, \cite{diener1999subjective} define subjective well-being as a general area of scientific interest, hence each specific construct related to it needs to be understood individually. One of these constructs is work well-being. It is discussed at length by \cite{schulte2010well}, who point to a positive relationship between work well-being and productivity at the societal level.

Very broadly, stress can be defined as a state of real or perceived disharmony that threatens homeostasis, i.e., a state in which equilibrium and optimal functioning, including body temperature of an organism, are threatened, by either intrinsic or extrinsic forces, i.e., stressors \citep{chrousos1992concepts}. Various physiological correlates to stress include blood pressure, heart rate, and galvanic skin response \citep{vrijkotte2000effects, schuler1980definition}. Prolonged stress can lead to cognitive impairments \citep{mcewen1995stress}, and neuronal disturbances resembling changes that are observed in the brain during depression \citep{de2005stress}.

A multitude of definitions for stress in organizational settings are collected and discussed by \cite{schuler1980definition}. The author concludes that these definitions ``suggest that individuals are 'under stress' particularly when the demands of the environment exceed (or threaten to exceed) a person's capabilities and resources to meet them or the needs of the person are not being supplied by the job environment.''.

In more recent times, the job demands-resources model \citep{karasek1990healthy, bakker2007job} is commonly used to explain employee well-being. The model generally divides job-related factors into two categories: demands and resources. Well-being is the outcome of the balance between these two categories, while job strain is produced by an imbalance between job resources and demands. Resources can be divided into personal and job resources. Personal resources are positive self-evaluations linked to resiliency and a sense of ability to control and impact upon the environment. On the other hand, job resources are physical, social, psychological and/or organizational aspects that are functional in achieving work goals, reducing demands, and stimulating personal growth \citep{xanthopoulou2009reciprocal}. Evidence suggesting job resources, personal resources, and work engagement are reciprocal over time, and support employee well-being exists \citep{xanthopoulou2009reciprocal}. Similarly, evidence of worker autonomy and social support increasing work engagement exists \citep{taipale2011work}. Work demands and continued job strain are connected to exhaustion and burnout \citep{demerouti2001job, xanthopoulou2007job}. Related to software development, the usage of information and communication technology is seen to be one possible source of stressors \citep{tarafdar2007impact}.

\subsection{Work Well-being and Emotions in Software Engineering}
\cite{sonnentag1994stressor} surveyed 180 software developers to identify factors related to burnouts, they discovered a lack of identification, (\ie, praise and recognition, and perceived pressures such as time pressure) to be related to stressors. Similar results have been obtained by \cite{singh2013health}, who surveyed Indian software developers and found mediating effects to stress with subjective well-being, social support, and meditation.
 
\cite{kuutila2020time} reviewed the effects of time pressure on software productivity and quality. The evidence shows lessened quality due to time pressure, while the evidence on productivity is two-fold: most cost and scheduling models assume increased total effort with compressed schedules, but empirical studies and experiments report increased efficiency under time pressure.

\cite{fucci2018need} investigated the effect of sleep deprivation on software developers and found that even a single night of sleep deprivation had a negative effect on software development quality. However, in a different study, it has been noted that two-thirds of developers work during normal working hours, while large differences between projects exist \citep{claes2018programmers}.

Interruptions and their effects on software development work have been investigated. \cite{tregubov2017impact} showed that developers working in multiple projects use a significant amount of their working time on context switching. \cite{sykes2011interruptions} discovered that senior developers and technical leads were experiencing more interruptions in their work in comparison to the regular staff at a software development company, guidelines on avoiding interruptions for software developers are also given in the work. \cite{brumby2019interruptions} has synthesized studies on interruptions' effects on productivity in software engineering to insights, some of which concern the types of interruptions. The most relevant insights to our study include ``Shorter interruptions are less disruptive than longer interruptions'' and ``Interruptions can cause stress, particularly e-mail interruptions.''.

Sentiment analysis has been defined as a series of methods, techniques, and tools for detecting and extracting subjective information, such as opinions and attitudes, from language \citep{liu2009handbook}. From software engineering context, \cite{jongeling2015choosing} compared and evaluated general sentiment analysis tools and their performance in the software engineering context, discovering that the tools evaluated did not agree with each other or manual labeling, thus concluding that tools for software development specific context are needed. 

There are a limited number of studies on the usage of emoticons by software developers, but \cite{claes2018use} have studied the use of emoticons by developers in two issue trackers. They found out project-level differences between Apache and Mozilla projects. Moreover, there were also differences between geographical locations, with developers from Europe and northern America using more emoticons.

With consideration of the pertinent literature, the novelty of our work lies in combining multiple data sources (experience sampling and repository mining), and examining the links between these data sources using multivariate models.

\subsection{Experience sampling method (ESM)}
\label{sec:ESM}
\subsubsection{Overview from Psychology}
The experience sampling method (ESM), also known as the daily diary method, studies everyday experiences and behavior in a natural environment, with data gathered both from both psychological and physiological sources \citep{alliger1993using}. The strengths of ESM lie in its empirical nature in which documentation of real-life experiences increase its ecological validity, its allowance of investigating within-person processes, its reduction of memory bias compared with other methods using self-reports, its allowance of investigating contingent behavior, and its ability to augment other research methods. Among possible weaknesses related to experience sampling are the self-selection bias, motivation issues in the acquired sample, the limited number of questions in data gathering, and the possible reactivity to the research setting \citep{scollon2009experience}. 

Experience sampling methods have been divided into three categories \citep{scollon2009experience} based on the time when the experiences are gathered: interval-contingent sampling, event-contingent sampling and signal-contingent sampling. Interval-contingent sampling refers to collecting data after a given time interval (e.g., hourly, daily, or weekly). In event-contingent sampling, data are gathered after specific events (e.g., after every meeting or social interaction). Lastly, signal-contingent sampling refers to a situation where participants in the study are prompted to answer at a randomly timed signal. A variety of devices can be used to remind subjects to respond to surveys and questionnaires, such as personal digital assistants, booklets, beepers, or wristwatches \citep{kimhy2006computerized}. However, reminders via email or SMS are also commonly applied. 

In previous studies on work well-being outside of software engineering, experience sampling methods and daily questionnaires have been used to study events, moods and behavior in a work setting. Some examples of the findings are that negative job events are five times more likely to be related to a negative mood than positive job events are to a positive mood \citep{miner2005experience}. Additionally, job satisfaction has been measured with experience sampling methodology and evidence has been found that affect and cognition are antecedents to job satisfaction \citep{ilies2004experience}. Continued cognitive engagement, more positive affect during work activity than during leisure activity and preference for work activities over leisure activities have been linked to workaholism in an ESM study \citep{snir2008workaholism}. Outside the work context, experience sampling has also been used to study interaction with information systems. For example, it has been found that an increase in the usage of Facebook predicted a lower life satisfaction level \citep{kross2013facebook}. The novelty of our work is to combine ESM data with data acquired from software repositories. 


\subsubsection{Challenges in Statistical Analysis}
Experience sampling methods produce time-series data that should be considered during analysis. As some statistical tests assume the independence of observations, non-independence in the time series data gathered with experience sampling is a problem needing action. \cite{west1991statistical} identify three main sources of non-independence that can occur in the data: auto-correlation, trend, and seasonality, all of which should be accounted for in an analysis.

Repeated measures over time can create auto-correlation, \ie, time-dependent data in violation of the assumption of independence. For example, the level of stress felt today is not completely independent on the level of stress felt yesterday. Controlling for the trend is important when cross-correlating time series, as underlying trends, create spurious correlations between the time series. For example, an increasing trend in the number of software engineers over time would create spurious correlations with many software engineering output measures such as commits and defect reports. The seasonality components usually refer to daily, weekly, monthly or yearly cycles, for example, stress levels could be perceived as stronger on Mondays.

\subsection{Negative results}
Publication bias ``is the tendency on the parts of investigators, reviewers, and editors to submit or accept manuscripts for publication based on the direction or strength of the study findings'' \citep{dickersin1990existence}. Publishing negative results have been seen to fight publication bias \citep{dirnagl2010fighting}. Still, evidence points toward decreased publishing of negative results in modern times \citep{fanelli2012negative}.

In software engineering there is also increased interest in allowing negative results to break the publication bias barrier. 
Related to our work, a couple of negative results have been published. Both roughly point out that neither general-purpose nor software engineering-specific sentiment analysis tools agree with manual labeling or with the results of each other in software engineering \citep{jongeling2017negative, lin2018sentiment}.


\section{Methodology}\label{sec:methodology}

An experience sampling study was conducted in a medium-sized software company in Finland. During our study, it employed four to five hundred people across its projects.

We developed a questionnaire that was sent to one of their teams developing a service with Agile methods and continuous delivery. Some elements from Scrum were present, the development process had iterations, after which results were presented, and future directions were planned in a retrospective. Tasks were tracked in a ``kanban'' board style with tickets: from there, they were completed as ready when they were deployed and tested to the staging environment. The project has a single customer and meetings with the customer were held almost weekly. The software is used in the daily operations of the customer, but it is not safety-critical software.

\subsection{Daily Questionnaire}
\label{sec:daily_questionnaire}

We constructed a short questionnaire, which was answered daily by the project team from the software company. The goal of this questionnaire was to produce data related to the work experience of the software project personnel, specifically developers. We piloted the questionnaire with the authors of this article. The aim was to produce a questionnaire that can be taken quickly, to achieve high response rates across a prolonged period. Therefore, we used single-item measurements, which have been shown in general to produce valid data in prior studies \citep{wanous1997overall, nagy2002using, elo2003validity}.

The questionnaire was constructed by picking relevant items from the survey done by \cite{elovainio2015stressful}, that studied work well-being of  physicians and was published in the \textit{Journal of Occupational Health Psychology}.  The questionnaire includes six single items that measure variables related to job well-being on an ordinal five-point scale. Thus, our questions represent theoretical concepts related to work health and well-being, as explained in section \ref{sec:background}. As the past survey was not done in the software engineering domain, we added one software engineering specific item to the questionnaire. Only one software-specific question was added to the questionnaire in order not to overload the respondents. This resulted in the following statements (without the emphasis) included in the questionnaire:
\begin{itemize}
\item \textit{I can make independent decisions in my work}. Individuals' \textbf{independence} and autonomy have been under study as a mediating factor between job demands and resources \citep{bakker2005job, xanthopoulou2007job}, \ie, there is evidence that increased autonomy in work tasks lessens the effects of job demands such as time pressure. 
\item \textit{I am in a hurry and have too little time to finish the task properly}. \textbf{Hurrying} to complete work, also known as time pressure, is a job demand, and has a complex relationship with performance \citep{bakker2007job, kuutila2020time}. It has been shown to be associated with increased performance in the short term \citep{nan2009impact,mantyla2014time}, but also higher stress \citep{svenson1993time} and even burnout \citep{donald2005work,bakker2005job,sonnentag1994stressor}.
\item \textit{I feel interrupted while working}. \textbf{Interruptions} to work increase the effort needed for task completion and have also been shown to increase time pressure and stress in the software development context \citep{mark2008cost}. The types of interruptions also play a role, with longer interruptions being worse for performance \citep{brumby2019interruptions}.
\item \textit{I experience ineffective software development (poor processes, poorly performing tools or poor communication with the development team)}. This question includes common topics related to productivity in software processes \citep{diaz1997software}, tools \citep{bruckhaus1996impact}, and communication \citep{wagner2018systematic}. 
\item \textit{I feel stressed (refers to a situation in which the respondent feels tense, restless, nervous, or anxious)}. In our case this refers to distress. \textbf{Stress} is modeled to be the result of an imbalance of demands and resources \citep{bakker2007job}, it has been linked to cognitive impairments \citep{mcewen1995stress}, and affective states related to depression \citep{de2005stress}.
\item \textit{I experience sleeping problems (difficulty in falling asleep or waking up several times during the night)}. \textbf{Problems sleeping} have been strongly linked to stress and increased job demands \citep{aakerstedt2002sleep, linton2004does}.
\end{itemize}

As previously stated, the questionnaire was constructed by picking relevant items from the survey done by \cite{elovainio2015stressful}. We did not opt for multiple items, that is multiple questions measuring the same variable. This is because the developers answering several dozens of questions daily would not have been practical nor possible. Our single items about \textbf{independence} and \textbf{interruptions} are from Karasek's Job Content Questionnaire \citep{karasek1998job}; the item measuring \textbf{Hurry} is from the Harris stress index \citep{harris1989nurse}; the item regarding stress is originally from the general health questionnaire ``GHQ-12'' \citep{goldberg1970psychiatric} and refers to distress; and lastly the question concerning sleeping problems is from the Jenkins scale \citep{jenkins1988scale}. These questions were slightly modified to fit our five-point scale: the respondents were asked to rate these six items with the question: ``How frequently has the following condition occurred since the last time you answered this survey?''. These items were then ranked on a five-point scale. From 1 to 5, the corresponding textual answers were ``very rarely or never'', ``rarely'', ``once in a while'', ``often'' and ``frequently or continuously''. Before starting the data collection, we met with the project personnel to explain the purpose of the study, and to clarify why daily answers were needed for the questionnaire.

The developed questionnaire was sent to the developers of the project over a period of 8 months (from April 10th, 2017 to January 12th, 2018). We used Webropol\footnote{\url{http://w3.webropol.com/start/}} to send the questionnaire every working day by email at 8 a.m. and to collect the responses. Developers who moved from or to another project, or started working in multiple projects at the same time, stopped answering the questionnaire. Developers with less than ten responses were discarded from the data analysis. 

For data analysis, a total of 526 responses were received from eight respondents. All responses included answers to all questions. None of the answers were preset, i.e. there was no pre-checked default answer. Developers could also simply not answer the questionnaire sent to them during some of the days. Multiple answers received during the same day by one individual were replaced with the mean of those answers, reducing the number of analyzable answers to 502. We also received another five answers during a weekend, and we removed these answers from analysis, further reducing the answers to 497. Considering the summer holidays, the total response rate is 37,5\% (526 / 1404) for eligible respondents. Looking at response times during the day before aggregating multiple answers, around 68,5\% were given between 7:00-10:00 a.m., and around 95\% during normal sliding working hours of 7:00-16:00. Two answers were given before 7:00 a.m., and a total of 19 after 5:00 p.m. The response rate was the highest during the first three months of the study (58\% of the total responses), decreasing steadily afterward with the last three full months having 23\% of total responses.

\subsection{Mining Software Repositories}

In Table~\ref{tab:var}, we provide the name and a short description of all the variables acquired from the software repositories. In the following subsections, we explain why and how these variables were acquired.

\begin{table}
\centering
\caption{Overview of software repository variables. All variables per day.}
\rotatebox{90}{
\label{tab:var}
\begin{tabular}{l|l|l}
\textbf{Variable} & \textbf{Description} & \textbf{Dimension}\\
\hline
  v1 commits & \#commits each developer made & Code Activity\\
  v2 loc & \# of code lines changed by the developer & Code Activity\\
  v3 filelogsum & \#files changed by the developer in all commits during the day & Code Activity \\
  \hdashline
  v4 nchat & \#chat messages sent & Amount of Communication \\
  v5 emoticon & \% of messages sent by the developer with an emoticon or an emoji & Expressed sentiment\\
  v6 joypr & \% of messages sent by the developer with joy emoticon or emoji & Expressed sentiment\\
  v7 sadconfusionsurprisepr & \% of messages sent by the developer with an emoticon or an emoji & Expressed sentiment\\
 & expressing sadness, confusion or surprise & Expressed sentiment\\
 \hdashline
  v8 negative valence & \% of messages sent by the developer with words with negative valence scores & Expressed sentiment\\
  v9 positive valence & \% of messages sent by the developer with words with positive valence scores & Expressed sentiment\\
  v10 minimum valence & The minimum score of the valence word for that day  & Expressed sentiment\\
  v11 maximum valence & The maximum score of the valence word for that day  & Expressed sentiment\\
  \hdashline
  v12 low arousal & \% of messages sent by the developer with words with low arousal scores & Expressed sentiment\\
  v13 high arousal & \% of messages sent by the developer with words with high arousal scores & Expressed sentiment\\
  v14 minimum arousal & The minimum score of the arousal word for that day & Expressed sentiment\\
  v15 maximum arousal & The maximum score of the arousal word for that day & Expressed sentiment\\
  \hdashline
  v16 meeting & binary variable of whether there was a meeting with the customer during the day & Job event\\
  v17 failure & \#times production tests failed during the day & Job event\\
\end{tabular}
}
\end{table}
\subsubsection{Version control system}

We used Perceval~\citep{duenas2018perceval} to extract the list of commits from the Git repository used by the project team. For each day of the period during which the developers answered the questionnaire, we computed for each respondent the number of commits made (\emph{ncommits}) and the number of lines of code modified (\emph{nloc}). While software development contains tasks not captured by these metrics, the number of commits and lines of code have been widely used as proxy measures for productivity in software engineering \citep{mockus2002two,boehm1981software}. Recent work has noted lines of code having the highest correlation with self-evaluated productivity \citep{murphy2019predicts}.

Entropy has been used to quantify the complexity of code changes in previous literature \citep{hassan2009predicting}. However, we decided to use the number of files changed by the developer each day, without considering the size of the project itself. This is because the number of developers grew during the project, some of whom did not answer the questionnaire. Result is the variable \textit{filelogsum}, which describes the number of times files were changed by the developer during the day, transformed to the base-10 logarithmic scale, as a result of the skewed nature of the distribution. 

\subsubsection{Mining chat messages}

Additionally, the company also provided us with a JSON dump of the chat room used by the developers. The specific tool used for communication changed during our study from Hipchat\footnote{\url{http://www.businessinsider.com/atlassian-launches-hipchat-successor-stride-2017-9}} to Slack\footnote{\url{https://slack.com/}}. From this chat archive, we computed the daily number of chat messages (\emph{nchat}) for each respondent.

We also translated lexicons used in the software engineering context for measuring arousal~\citep{mantyla2017bootstrapping} and valence to Finnish to do rudimentary sentiment analysis on the chat logs. Chat logs were lemmatized using the open-source software Voikko~\citep{Voikko}, and then scored on valence and arousal using the translated lexicons. The arousal or valence scores in the lexicons range from 1 to 9, and thus are centered around 5. Hence low valence and arousal are shown in scores under 5, and high valence and arousal in scores over 5. We use this information in the variables \emph{negative valence}, \emph{positive valence}, \emph{low arousal} and \emph{high arousal}. Hence, the variable \emph{negative valence} contains the percentage of messages containing at least one word with a valence score below 5 and the variable \emph{positive valence} denoted as the percentage of messages containing at least one word with a valence score above 5. The same method was applied to for variables \emph{low arousal} and \emph{high arousal}. Similarly, we also calculated the maximum and minimum arousal and valence scores for each day for each developer, and these are found in the variables \emph{minimum valence}, \emph{maximum valence}, \emph{minimum arousal}, and \emph{maximum arousal}.

We also extracted emoticons and emojis that were used in the chat messages. Emoticons are textual representations of human emotion using only keyboard characters such as letters, numbers, or punctuation marks. Emojis refer to "picture characters" or pictographs~\citep{miller2016blissfully}. Similar to some of the authors' previous work \citep{claes2018use}, we manually classified the emoticons to the basic emotions of Plutchik's wheel of emotions \citep{plutchik1991emotions}: joy, sadness, surprise, confusion, and anger. The used list of emoticons and emoji, and their associated emotions, is available online\footnote{\url{https://github.com/M3SOulu/autotime-esm}}. The first and third authors classified the emoticons and achieved a 79.5\% agreement with a Cohen's kappa of 0.7, after which we went through the cases where we disagreed. With these emoticons, we calculated the percentage of messages containing \emph{emoticons} and emojis, the percentage of messages containing emoticons and emoji \emph{related to joy}, and the percentage of messages containing emoticons and emojis \emph{related to surprise, sadness, and confusion}. Due to the low number of emoticons and emoji for the latter group of emotions, we combined them in one variable named \emph{sadconfusionsurprise-emo}. For conciseness in the results section of this manuscript, emoticons refer to both emoticons and emojis.

\subsubsection{Factor Analysis and Measurement Model}

We used factor analysis to study the structure of the underlying variables in our data set \citep{thompson2004exploratory}.  We explored the data sources from Table \ref{tab:var} with the \emph{fa.parallel} function\footnote{\url{https://www.rdocumentation.org/packages/psych/versions/2.0.8/topics/fa.parallel}} for the optimum number of factor, then we used the \emph{fa} function\footnote{\url{https://www.rdocumentation.org/packages/psych/versions/2.0.8/topics/fa}} to find the minimum residual (minres) solution using 100 iterations. The resulting factors are in the left side of Figure \ref{fig:image}. 

For these factors we computed the goodness of fit statistics, which shows a very good fit (Table \ref{tab:gfit}).
For choosing the goodness of fit statistics, we followed the figure given by \cite{sun2005assessing} and give sample based goodness of fit indices Tucker Lewis index~\citep{tucker1973reliability} and the root mean square residual (RMSR). The TLI or Non-Normed Fit Index is a fit measure comparing the fit in relation to the null model \citep{marsh1996evaluation}. RMSR is a descriptive fit defined as ``is defined as the square root of the mean of the squared fitted residuals'' \citep{schermelleh2003evaluating} . 
Measures of TLI and RMSR indicate a very good fit. While unusual, the TLI can have values greater than one, see discussions by \cite{anderson1984effect} and by \cite{UnusualTLIvalues}.




\begin{table}[t]
\centering
\caption{Goodness of fit statistics for factors discovered with exploratory factor analysis.}
\label{tab:gfit}
\begin{tabular}{l|l|l|l|l|l|l|l}
Factor & TLI & RMSR \\
\hline
Productivity & 0.99 & 0.01 \\
Positive Valence  & 1.029 & 0.01 \\
Negative Valence  & 1.01 & 0 \\
High Arousal  & 1.01 & 0 \\
Low Arousal  & 1.01 & 0 \\
Joyemo  & 1.01 & 0 \\
\end{tabular}
\end{table}

Our measurement model (Fig.~\ref{fig:image}) shows the relationships between latent variables and their indicators \citep{BOLLEN20017282}. On the left factors created by exploratory factor analysis are shown, on the right correlations between variables acquired with the questionnaire are shown. The oval shapes under ``Repositories'' denote the factors from Table~\ref{tab:gfit}. The rectangles under ``Repositories'' show the variables from Table~\ref{tab:var}. Lines between variables and factors show weights, with dotted lines signaling negative weights. On the 

\subsubsection{Generalized Linear Mixed Effect Models}\label{sec:lmem}
We used generalized linear mixed effects models as they can be used to study both fixed and random effects. We used the package nlme \citep{pinheiro2017package} to construct the models because it can easily accommodate auto-correlation structures. The variables specified in our measurement model were evaluated as fixed effects. For random effects, we used a unique respondent identifier, variable specifying the day of the week (``weekday''), and a time variable designating the day during the study (i.e. the first day of the study as 1, the second as 2, and so on). We used the function r.squaredGLMM from the package MuMIn \citep{barton2009mumin} to calculate both the marginal and the conditional $R^2$ values. Marginal $R^2$ values represent the variance explained by the fixed effects, while the conditional $R^2$ values are interpreted as a variance explained by the entire model, including both fixed and random effects. Calculating marginal and conditional $R^2$ values for mixed effects models is based on the work of \cite{nakagawa2013general}. When constructing models for individuals, we took the four respondents with the highest number of answers to the questionnaire.

\subsubsection{Seasonality and auto-correlation}
We studied the trends and seasonality in our collected data with the R function \emph{decompose} \footnote{https://www.rdocumentation.org/packages/stats/versions/3.6.2/topics/decompose} and found weekly seasonality for all the software repository variables. The weekly seasonality of the chat messages is the highest. The average number of chat messages sent on Mondays were 30.7, 25.8 on Tuesdays, 30.8 on Wednesdays, 41.6 on Thursdays, and 45.7 on Fridays. By comparison, the seasonality of commits per day is weaker. The average commits on Mondays was 9.7, 7.8 on Tuesdays, 7.8 on Wednesdays, 11.2 on Thursdays, and 8.5 on Fridays. To account for time-series data and control for weekly seasonality in the data, we added a weekday variable as a random effect to the models. 

We also investigated the auto-correlations of the data with the \emph{acf} function\footnote{\url{https://www.rdocumentation.org/packages/forecast/versions/8.3/topics/Acf}} of the forecast R package \citep{hyndman2007automatic}, and found strong auto-correlations for all the questionnaire variables. As a consequence, we added these variables as random effects to our generalized linear mixed effects of models. In practice, this means we used the corarma function\footnote{\url{https://www.rdocumentation.org/packages/nlme/versions/3.1-137/topics/corARMA}} with a 10 day moving average structure when making general models, and a 5 day moving averages when creating models for individuals. We observed no meaningful differences between results in our models based on the used auto-correlation structure whether it was a one day average or different moving averages, but we had troubles with model convergence depending on the used auto-correlation structure. Convergence problems are related to the complexity of random effects, and are further discussed in section \ref{sec:threats}.

\begin{figure*}[htp]
\centering
\caption{The measurement model resulting from factor analysis, showing variable weights to factors and correlations between latent variables used in our study.}
\includegraphics[width=\textwidth]{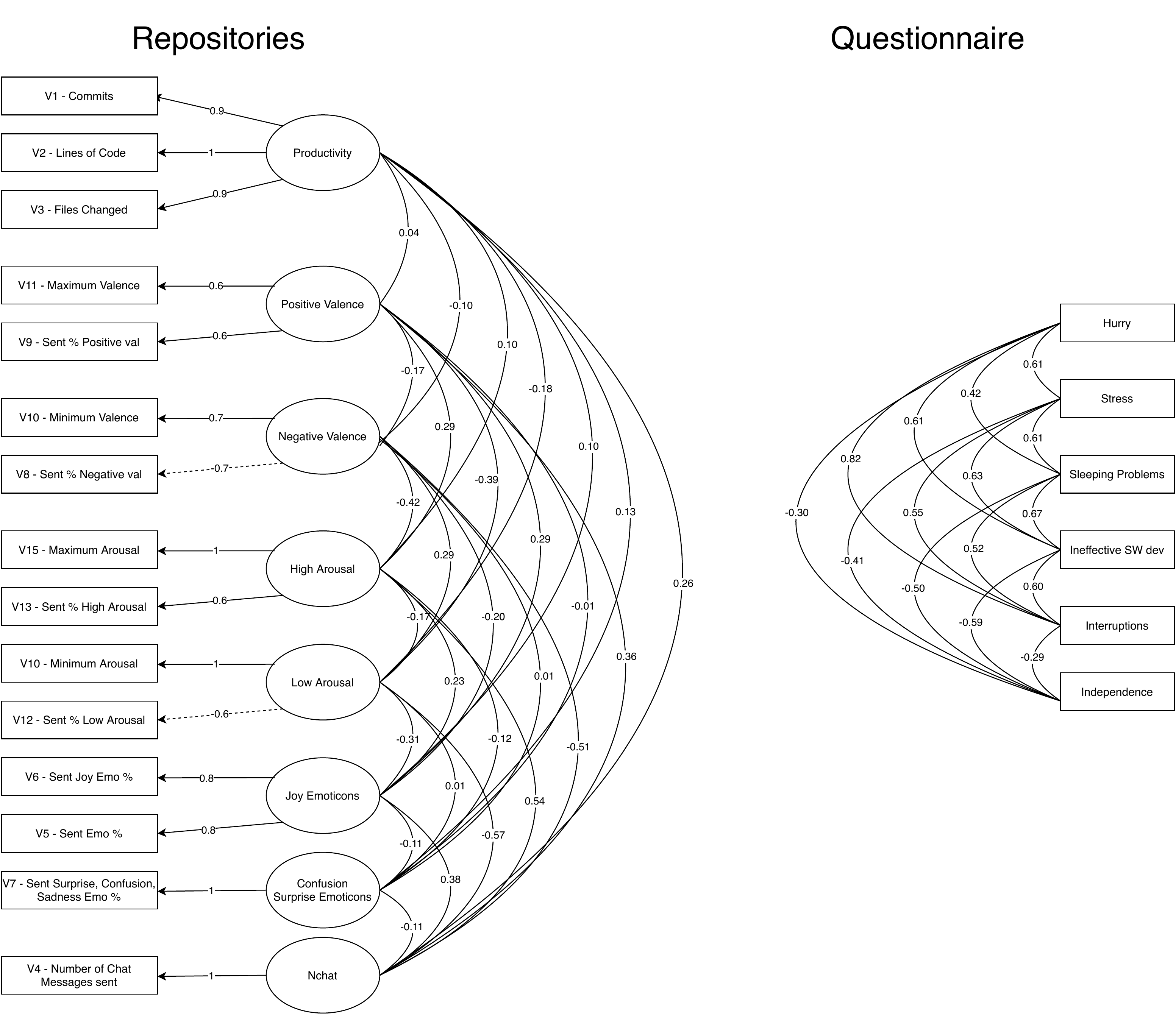}
\label{fig:image}
\end{figure*}








\subsection{Semi-structured Interviews}\label{sec:methodologyinterviews}

Semi-structured interviews have for long been advocated for and used to collect qualitative data about phenomena in software engineering \citep{hove2005experiences}. Semi-structured interviews have been recommended as a supplement to surveys and questionnaires, when important questions remain after collecting quantitative data \citep{adams2015conducting}. Hence, after publishing our previous work \citep{kuutila2018using}, from which this paper is based, we conducted semi-structured interviews to answer ``why questions'' about the quantitative data we had gathered. The interview questions were thus designed to better explain our prior results, and the translations of these questions are available at Github\footnote{\url{https://github.com/M3SOulu/autotime-esm}}.

For conducting and designing the interviews, we followed the guidelines given by ~\cite{adams2015conducting}. Drafting interview questions and making the interview guide was done collectively by the authors, aiming for open-ended questions for which we could ask follow-up questions and query clarifications. Since visualizing timelines and results have been advocated for project-level retrospectives~\cite{bjarnason2014reflecting}, in our case, we decided to help remembrance and recollection by visualizing the answers to the questionnaire by sending graphs of individual-level responses to the interviewees before the interview. The first author interviewed the project manager and two developers for this study. One interview was done in person and two over a video call. The interviews totaling almost three hours were recorded with written permission from the interviewees and transcribed verbatim after the interview process.
 
We present some background information on the interviewees in Table~\ref{tab:r5}. The project manager was not part of the data collection of the quantitative research questions, but we believed project managers' views on the project were valuable. 

\begin{table}[t]
\centering
\caption{Interviewee characteristics.}
\label{tab:r5}
\begin{tabular}{l|l|l|l|l|l|l|l}
ID Role & \#Years in SW industry  & Degree  \\
\hline
PM/Project Manager & 13 years & Masters \\
Developer 1  & 7 years  & Masters \\
Developer 2  & 4 years  & Masters \\
\end{tabular}
\end{table}
  
The interview started with questions about the questionnaire procedure itself, and about the individual level graphs we mentioned previously . The main aim of this was to help the respondent recall the answering period, but also to get any recommendations to make the procedure easier and to increase the response percentage. After this, we asked a question related to the results of the previous conference paper \cite{kuutila2018using}, and the kind of explanations the respondents might have. Next, we asked questions related to the new variables we were bringing to the analysis: expressed sentiment, emoticon usage, and job events. Related to job events, we asked for any recollections of hurry and stress periods during the questionnaire.

We followed the analytical strategy of \cite{schmidt2004analysis} for analyzing the transcripts of the semi-structured interviews. As advocated by \cite{schmidt2004analysis}, the analysis process comprised repeated and intensive reading of the transcriptions and the development of a coding scheme from analytical categories. In the end, our very simple scheme contained three different codes: (1) facilitating activities helping well-being; (2) barriers to well-being and (3), explanations of our results. The fourth step of quantifying the material involved mainly finding codes that were uniform and coherent across the interviewees. These have been mentioned in the results in the form of whether the respondent agreed or disagreed on topics, and emphasizing themes that were uniform across the three interviews. Not mentioning a particular topic was not interpreted as a disagreement.



\section{Results}\label{sec:results}
\subsection{RQ1 - \RQOne}

Our main motivation for this research question was to understand how well-being was felt in the software project under study. In particular, if several individuals on the development team reported similar well-being and affective states simultaneously during the development project, it could be that external demands such as deadlines could affect the whole development team at the same time. For example, there is evidence that part of work-related stress is shared within organizations \citep{semmer1996shared}. Additionally, related work on time pressure has called for organizational-level studies  \citep{silla2014shared}. 

The values produced by Krippendorff’s alpha \citep{krippendorff2011computing} are between 1 (perfect agreement), 0 (statistically unrelated), and -1 (perfect disagreement). To interpret the values, Krippendorff proposed thresholds \citep{krippendorff1980reliability}, where a value of 0.2 is considered poor, and values greater than 0.7 are good. \textbf{We observe poor disagreement between respondents for all questionnaire variables.} Table~\ref{tab:agreement} shows values from -0.214 for Sleeping Problems up to -0.099 for Hurry. These negative values could imply two things: either the respondents feel each affective state individually rather than on a group level, or they use different calibrations of the scales. That is, some individual respondents consider a value of 2 for Hurry normal while others consider it to be exceptional.

\begin{table}[t]
\centering
\caption{Inter-coder agreement of the respondents}
\label{tab:agreement}
\begin{tabular}{ll}
\textbf{Variable:}    		& \textbf{Krippendorff's Alpha:} \\
Hurry                            & -0.0993              \\
Independence                     & -0.119               \\
Ineffective Software Development & -0.178               \\
Interruptions                    & -0.161               \\
Sleeping Problems                & -0.214               \\
Stress                           & -0.155              
\end{tabular}
\end{table}

\begin{table}[htp]
\caption{Generalized linear mixed models predicting questionnaire variables with the previous working days repository variables. A p-value of 0.05 or less is denoted in \textbf{bold}.}
\label{tab:t1}
\begin{tabular}{p{9em}|p{4em}|p{4em}|p{4em}|p{4em}|p{4em}|p{4em}}
Predicted: & Hurry & Stress & Sleep & Inter- & Ineffe& Indepe\\
 &  &  &  & ruptions & ctive & ndence \\ 
  \hline
prod & 0.07 (\textbf{0.02}) & 0.19 (\textbf{<0.001}) & 0.06 (0.28) & -0.02 (0.71) & 0.02 (0.47) & -0.03 (0.20) \\ 
\hdashline
nchat & -0.03 (0.34) & -0.06 (0.27) & 0.05 (0.33) & -0.01 (0.89) & -0.02 (0.62) & 0.01 (0.89) \\ 
\hdashline
pval & -0.03 (0.29) & 0.04 (0.57) & 0.08 (0.20) & -0.08 (0.14) & 0.02 (0.67) & 0.10 (\textbf{<0.001}) \\ 
nval & 0.03 (0.32) & -0.03 (0.58) & -0.09 (0.15) & 0.03 (0.52) & 0.04 (0.31) & 0.04 (0.09) \\  
\hdashline
har & 0.02 (0.64) & 0.02 (0.77) & 0.02 (0.77) & 0.13 (\textbf{0.01}) & 0.09 (\textbf{0.03}) & -0.04 (0.11) \\ 
lar & -0.06 (0.10) & <-0.01 (0.93) & -0.04 (0.49) & -0.05 (0.23) & -0.01 (0.71) & -0.02 (0.48) \\
\hdashline
joyemo & 0.01 (0.79) & 0.03 (0.26) & -0.03 (0.42) & -0.02 (0.52) & -0.01 (0.53) & 0.02 (0.22) \\ 
scsemo & -0.05 (\textbf{0.05}) & 0.44 (0.25) & -0.20 (0.61) & 0.28 (0.38) & 0.31 (0.18) & 0.35 (\textbf{0.03}) \\ 
\hdashline
failure event & -0.06 (0.07) & 0.03 (0.56) & 0.01 (0.91) & -0.04 (0.38) & <-0.01 (0.95) & -0.02 (0.42) \\ 
meeting & -0.13 (\textbf{0.001}) & 0.01 (0.92) & <0.01 (0.97) & -0.01 (0.88) & 0.04 (0.29) & -0.04 (0.13) \\ 
   \hline
Random effects residual stddev : &  &  &  &  &  &  \\ 
respondent & 0.61 & 0.85 & 0.77 & 0.71 & 0.50 & 0.41 \\ 
weekday& 0.00 & 0.00 & 0.00 & >0.01 & 0.00 & >0.01 \\ 
date& 0.00 & 0.00 & 0.01 & >0.01 & 0.05 & >0.01 \\ 
\hline
Marginal $R^2$ & 0.01 & 0.02 & 0.01 & 0.01 & <0.01 & 0.01 \\ 
Conditional $R^2$ & 0.76 & 0.39 & 0.44 & 0.67 & 0.83 & 0.68 \\ 
\hline
Null Model C. $R^2$ & 0.74 & 0.39 & 0.44 & 0.67 & 0.80 & 0.66 \\ 
\end{tabular}
\end{table}

\subsection{RQ2 - \RQTwo}\label{sec:rsc2}

As developers' responses to the questionnaire differed at the same time points, we analyzed them in relation to several factors derived from software repositories with a one-day time lag while taking the individual into account. For analysis, we chose generalized linear mixed models, and we use all the predictors due to the exploratory nature of our study. In Tables~\ref{tab:t1} to~\ref{tab:ind} we investigate the relationship from software repositories to our questionnaire responses with a one-day lag, i.e. can the previous days repository metrics predict current days questionnaire responses. As the questionnaire was sent each morning and most of the responses were also given in the morning (see the end of Section ~\ref{sec:daily_questionnaire}), using previous days' repository data seemed most reasonable to us.

In generalized linear mixed effect models (Section \ref{sec:lmem}), the marginal $R^2$ value represents the variance explained by the fixed effects, while the conditional $R^2$ value is interpreted as a variance explained by the entire model, including both fixed and random effects. We also provide a null model conditional $R^2$ value. They refer to models where only random variables are used to explain the predicted variable. Random variables in our case were the respondent ID, and the auto-correlation variables weekday and date as a number from 1 to 240.

Table~\ref{tab:t1} shows the models predicting questionnaire answers with the previous working days repository variables. In other words, do the actions of the previous working day predict well-being and the answer on the questionnaire to be completed the following morning. As we can see in Table~\ref{tab:t1} the conditional $R^2$ value is considerably higher than the marginal $R^2$ value in every model, meaning random effects explain overwhelmingly more variance in the predicted variable than fixed effects. When looking inside random effects, the respondent ID variable is mostly behind the dominant conditional $R^2$ value. This means that the individual in question has the highest effect on the prediction of the questionnaire variable. 

The highest effect of fixed effects can be found for prediction stress, with a marginal $R^2$ value of 0.02. For the fixed effects in the other models predicting questionnaire variables, a marginal $R^2$ value of 0.01 or less is found by the models. 

Although the marginal $R^2$ value is small, we go through the statistically significant regression coefficients as they may spark future works. When looking at the predictors in fixed effects, the highest coefficient is in productivity when predicting stress (p < 0.001). Productivity was also a significant predictor of a hurry. In other words, the developer's previous day's higher productivity was associated with experiencing hurry and stress the next day. We also found that expressing positive valence (pval) was associated with increased independence the next day but so was using sad or confused emoticons (scsemo). Therefore, it may be that independence is increased in both expressing positive and negative emotions. Expressing elevated arousal (har) was associated with developers reporting more interruptions and ineffective software development. Finally, we find that meetings reduced the feeling of hurry the next day.

Because the effect of the respondent was very high when predicting questionnaire outcomes, as seen in Table~\ref{tab:t1}, we constructed models with the data from four the individuals with the highest number of responses to the questionnaire. Tables~\ref{tab:hur} to~\ref{tab:ind} show models predicting questionnaire answers with a single individual's data. Similar to Table~\ref{tab:t1}, the questionnaire answers were predicted using the previous day's repository variables. Empty columns in the tables mean that the model did not converge. We also gave a null model conditional $R^2$ value, the amount of variance predicted solely by random effects, which in individual models solely contain the weekday and the date as the number of days from the start of the study period. 

Depending on the individual, predicting questionnaire answers with variables related to software repositories can achieve a marginal value of $R^2$ up to 0.26. This is in harsh contrast to the general model in Table~\ref{tab:t1}, where the marginal $R^2$ value did not exceed 0.02. 

Table~\ref{tab:hur} shows models for predicting hurry for three individuals, with the $R^2$ value varying between 0.10 and 0.26. Comparing the general model and the individual models with respect to hurry reveals the following. Productivity is associated with reduced hurry for developer B which it opposed the general model. Such oppositing results between developers explain the low $R^2$ value in the general model. 


Table~\ref{tab:stress} has two models, as for two individuals the model did not converge. For the two developers, the marginal $R^2$ values were at 0.10 and 0.09 respectively. For developer A, a p-value of 0.01 was calculated for productivity with a positive coefficient. For developer C, a negative coefficient and a p-value of 0.02 was calculated with a number of chat messages. Productivity also has a positive value for developer C and can also be found in the general model in Table~\ref{tab:t1} as a significant predictor. The number of chat messages was also negative for developer A, but it cannot be found in the general model. Other predictors that have the same sign for the two individuals are failure events and meetings.

Table~\ref{tab:sleep} shows four models depicting the prediction of sleeping problems by the respondents. The marginal $R^2$ values vary between 0.10 and 0.24. Significant predictors were high arousal for developer A with a positive relationship and  p-value of 0.02. For developer Bm, the significant predictor was meetings with a negative relationship and a p-value of 0.05. Lastly, for developer C, the predictor was joy emoticons and emoji with a p-value of 0.04 and a negative relationship. None of these predictors were in the general model in Table~\ref{tab:t1}. Uniform signs across the developers' could be found for negative valence in a negative relationship with sleeping problems.

Generally, individual models for the prediction of interruptions and ineffective software development achieve lower marginal $R^2$ values compared to the other three questionnaire questions, with only one model achieving a marginal $R^2$ value of 0.1. No statistically significant predictors could be found for ineffective software development, but two could be found for interruptions in Table~\ref{tab:inter}. These are productivity for developer B with a negative relationship and a p-value of 0.02, and for developer C low arousal with a negative relationship and a p-value of 0.04. None of these was a significant predictor in the general model in Table~\ref{tab:t1}. Lastly, Table~\ref{tab:ind} shows models for individuals predicting independence for one individual. The marginal $R^2$ value is 0.13. There were no statistically significant predictors.


\textbf{ Summary: We found no general model to predict software developer's well-being from software repositories. Yet, it seems that the well-being of each individual has different predictors.}  



\begin{table}[htp]
\caption{Generalized linear mixed models predicting hurry for the next day with today's repository variables. A p-value of 0.05 or less is denoted in \textbf{bold}.}
\label{tab:hur}
\begin{tabular}{p{8em}|p{7em}|p{7em}|p{7em}|p{7em}}
Variable: & DevA & DevB & DevC & DevD\\ 
\hline
prod & 0.11 (0.07) &  -0.53 (\textbf{<0.001}) & 0.08 (0.47) &    \\
\hdashline
nchat & -0.02 (0.80) & 0.15 (0.23)  & -0.09 (0.51) &    \\
\hdashline
pval & -0.06 (0.56) & -0.51 (\textbf{<0.001})  & -0.09 (0.64) &    \\
\hdashline
nval & -0.06 (0.43) & -0.19 (0.24)  & -0.11 (0.44) &    \\
\hdashline
har & 0.12 (0.25) & 0.06 (0.61)  & -0.28 (0.14) &    \\
\hdashline
lar & -0.02 (0.74) & -0.04 (0.71)  & -0.12 (0.41) &    \\
\hdashline
joyemo & -0.06 (0.08) &  0.02 (0.71) & 0.20 (\textbf{0.03}) &    \\
\hdashline
scsemo & 0.37 (0.30) & -0.99 (0.18)  & -0.08 (0.93) &    \\
\hdashline
meeting & -0.03 (0.74) & -0.22 (0.24)  & -0.02 (0.90) &    \\
\hdashline
failure event & -0.11 (0.08) & -0.05 (0.68) &   -0.08 (0.47) &    \\
\hline
Marginal $R^2$ & 0.11 &  0.26 & 0.10 &    \\
Conditional $R^2$ & 0.11 & 0.26  &  0.10 &    \\
\hline
Null Model C. $R^2$ & <0.01 & 0  & 0 &  \\
\end{tabular}
\end{table}

\begin{table}[htp]
\caption{Generalized linear mixed models predicting stress for the next day with today's repository variables. A p-value of 0.05 or less is denoted in \textbf{bold}.}
\label{tab:stress}
\begin{tabular}{p{8em}|p{6em}|p{6em}|p{6em}|p{6em}}
Variable: & DevA & DevB & DevC & DevD\\ 
\hline
prod & 0.38 (\textbf{0.01}) &   & 0.15 (0.23) &    \\
\hdashline
nchat & -0.25 (0.16) &   & -0.39 (\textbf{0.02}) &    \\
\hdashline
pval & -0.13 (0.62) &   & 0.17 (0.49) &    \\
\hdashline
nval & -0.19 (0.28) &   & 0.19 (0.32) &    \\
\hdashline
har & 0.24 (0.34) &   & -0.30 (0.19) &    \\
\hdashline
lar & -0.18 (0.31) &   & 0.08 (0.64) &    \\
\hdashline
joyemo & 0.02 (0.75) &   & -0.03 (0.81) &    \\
\hdashline
scsemo & 0.97 (0.30) &   & -0.14 (0.91 &    \\
\hdashline
meeting & -0.03 (0.89) &   &  -0.13 (0.53) &    \\
\hdashline
failure event & 0.25 (0.10) &   & 0.09 (0.55) &    \\
\hline
Marginal $R^2$ & 0.10 &   & 0.09 &    \\
Conditional $R^2$ & 0.10 &   & 0.09 &    \\
\hline
Null Model C. $R^2$ & <0.01 &   & 0 &    \\
\end{tabular}
\end{table}

\begin{table}[htp]
\caption{Generalized linear mixed models predicting sleeping problems for the next day with today's repository variables. A p-value of 0.05 or less is denoted in \textbf{bold}.}
\label{tab:sleep}
\begin{tabular}{p{8em}|p{6em}|p{6em}|p{6em}|p{6em}}
Variable: & DevA & DevB & DevC & DevD\\ 
\hline
prod & 0.19 (0.11) &  -0.41 (0.07) & -0.03 (0.77) &  -0.30 (0.57)  \\
\hdashline
nchat & <-0.01 (0.99) & 0.29 (0.13)  & -0.19 (0.13) &  0.05 (0.72)  \\
\hdashline
pval & -0.06 (0.80) & -0.28 (0.11)  & 0.10 (0.55) & -0.04 (0.84)   \\
\hdashline
nval & -0.23 (0.16) &  -0.05 (0.82) & -0.14 (0.31) &  -0.17 (0.35)  \\
\hdashline
har & 0.54 (\textbf{0.02}) &  0.04 (0.83) & -0.16 (0.38) & 0.07 (0.73)   \\
\hdashline
lar & -0.03 (0.84) &  -0.12 (0.48) & 0.09 (0.51) &  0.27 (0.19)  \\
\hdashline
joyemo & -0.04 (0.63) &  -0.05 (0.51) & -0.18 (\textbf{0.04}) &  0.32 (0.44)  \\
\hdashline
scsemo & 0.30 (0.70) & -0.04 (0.97)  & -0.13 (0.88) &  -1.6 (0.75)  \\
\hdashline
meeting & 0.20 (0.20) & -0.49 (\textbf{0.05})  & 0.01 (0.98) & 0.20 (0.44)   \\
\hdashline
failure event & -0.03 (0.84) & 0.01 (0.95)  & 0.01 (0.91) &  0.03 (0.77)  \\
\hline
Marginal $R^2$ & 0.11 & 0.14  & 0.10 &  0.24  \\
Conditional $R^2$ & 0.11 & 0.16  & 0.10 &  0.24  \\
\hline
Null Model C. $R^2$ & 0 & 0  & 0 &  0  \\
\hline
\end{tabular}
\end{table}

\begin{table}[htp]
\caption{Generalized linear mixed models predicting interruptions for the next day with today's repository variables. A p-value of 0.05 or less is denoted in \textbf{bold}.}
\label{tab:inter}
\begin{tabular}{p{8em}|p{6em}|p{6em}|p{6em}|p{6em}}
Variable: & DevA & DevB & DevC & DevD\\ 
\hline
prod & 0.06 (0.30) &  -0.45 (\textbf{0.02}) & -0.09 (0.42) &    \\
\hdashline
nchat & <-0.01 (0.99) & -0.09 (0.60)  & -0.06 (0.68) &    \\
\hdashline
pval & -0.11 (0.28) & -0.19 (0.27)  & -0.03 (0.88) &    \\
\hdashline
nval & 0.08 (0.28) &  0.03 (0.90) & -0.03 (0.84) &    \\
\hdashline
har & 0.03 (0.77) &  0.15 (0.39) & 0.15 (0.44) &    \\
\hdashline
lar & 0.03 (0.60) &  -0.12 (0.41) & -0.31 (\textbf{0.04}) &    \\
\hdashline
joyemo & -0.04 (0.25) & -0.01 (0.92) & 0.01 (0.97) &    \\
\hdashline
scsemo & 0.15 (0.66) &  -0.61 (0.56) & 1.5 (0.17) &    \\
\hdashline
meeting & 0.14 (0.13) &  -0.10 (0.68) & 0.11 (0.45) &    \\
\hdashline
failure event & -0.01 (0.83) &  <-0.01 (0.99) & -0.23 (0.06) &    \\
\hline
Marginal $R^2$ & 0.06 & 0.10  & 0.09 &    \\
Conditional $R^2$& 0.06 &  0.10 & 0.09 &    \\
\hline
Null Model C. $R^2$ & 0 &  0 & 0 &    \\
\end{tabular}
\end{table}

\begin{table}[htp]
\caption{Generalized linear mixed models predicting ineffective software development for the next day with today's repository variables. A p-value of 0.05 or less is denoted in \textbf{bold}.}
\label{tab:inef}
\begin{tabular}{p{8em}|p{6em}|p{6em}|p{6em}|p{6em}}
Variable: & DevA & DevB & DevC & DevD\\ 
\hline
prod & 0.02 (0.43) &  -0.07 (0.64) & <-0.01 (0.99) &    \\
\hdashline
nchat & <0.01 (0.95) & 0.02 (0.86)  & 0.02 (0.85) &    \\
\hdashline
pval & -0.01 (0.89) & -0.08 (0.54)  & 0.02 (0.87) &    \\
\hdashline
nval & 0.07 (0.06) & -0.09 (0.58)  & -0.13 (0.26) &    \\
\hdashline
har & -0.05 (0.32) &  0.22 (0.12) & 0.15 (0.30)  &    \\
\hdashline
lar & 0.01 (0.85) & 0.01 (0.92)  & <-0.01 (0.98) &    \\
\hdashline
joyemo & <-0.01 (0.90) & -0.03 (0.62)  & -0.02 (0.83) &    \\
\hdashline
scsemo & 0.07 (0.70) & -0.05 (0.95) &  -0.02 (0.97)  &    \\
\hdashline
meeting & 0.05 (0.32) & -0.10 (0.63)  & 0.10 (0.38) &    \\
\hdashline
failure event & 0.01 (0.65) & -0.02 (0.63)  & -0.11 (0.24) &    \\
\hline
Marginal $R^2$ & 0.02 &  0.06 & 0.02 &    \\
Conditional $R^2$ & 0.02 &  0.06 & 0.02 &    \\
\hline
Null Model C. $R^2$ & 0 &  0 &  0&    \\
\end{tabular}
\end{table}

\begin{table}[htp]
\caption{Generalized linear mixed models predicting independence for the next day with today's repository variables. A p-value of 0.05 or less is denoted in \textbf{bold}.}
\label{tab:ind}
\begin{tabular}{p{8em}|p{6em}|p{6em}|p{6em}|p{6em}}
Variable: & DevA & DevB & DevC & DevD\\ 
\hline
prod &  &   & -0.01 (0.94) &    \\
\hdashline
nchat &  &   & -0.16 (0.16) &    \\
\hdashline
pval &  &   & 0.20 (0.23) &    \\
\hdashline
nval &  &   & 0.23 (0.08) &    \\
\hdashline
har &  &   & 0.12 (0.42) &    \\
\hdashline
lar &  &   & -0.19 (0.11) &    \\
\hdashline
joyemo &  &   & -0.02 (0.84) &    \\
\hdashline
scsemo &  &   & -0.33 (0.70) &    \\
\hdashline
meeting &  &   & 0.14 (0.21) &    \\
\hdashline
failure event &  &   & -0.09 (0.38) &    \\
\hline
Marginal $R^2$ &  &   & 0.13 &    \\
Conditional $R^2$ &  &   & 0.13 &    \\
\hline
Null Model C. $R^2$ &  &   & 0 &    \\
\end{tabular}
\end{table}




\subsection{RQ3 - \RQThree}\label{sec:rsc3}

In RQ3, we examined whether developer productivity measured as a factor can be predicted with all of the other factors of our model, that is both the remaining software repository variables, as well as the questionnaire answers. The productivity factor consists of the number of commits, lines of code, and the number of files changed (Fig. ~\ref{fig:image}).

Table~\ref{tab:rq3} shows five different models for predicting productivity, one made using all the data and four made using individual developers with the most answers to the questionnaire, similar to RQ2. The $R^2$ values for the fixed effects show that again, random effects explain more than fixed effects, with marginal $R^2$ value of 0.03 and conditional $R^2$ value of 0.52. Again, the random effects refer to control variables, which are used to explain the predicted variable, that is the respondent ID, the day of the week, and the date as a number from the start of the study, with the first day being one.

The three models showing individuals in Table~\ref{tab:rq3} show individual variability, as only one predictor is statistically significant for one developer. Predictors with the same sign for all three individuals are the number of chat messages, negative valence, failure events, and independence. We can also see that the marginal $R^2$ value rises from 0.03 of the general model to 0.05-0.20 depending on the individual.

\textbf{
This result is highly similar to what we observed in RQ2. To summarize, how experienced well-being and actions predict productivity significantly vary between individuals. 
}


\begin{table}[htp]
\caption{Generalized linear mixed models predicting productivity during the same day. The general model and the four different individuals' models. A p-values 0.05 or less is denoted in \textbf{bold}.}
\label{tab:rq3}
\begin{tabular}{p{9em}|p{5em}|p{5em}|p{5em}|p{5em}|p{5em}}
Variable: & All & DevA & DevB & DevC & DevD\\ 
\hline
nchat & 0.05 (0.35) &  0.04 (0.73) & 0.15 (0.15) &  0.19 (0.19)  \\
\hdashline
pval & 0.02 (0.74) & -0.02 (0.91)  & -0.08 (0.42) &   0.06 (0.78) \\
\hdashline
nval & -0.05 (0.48) & -0.10 (0.51)  & -0.05 (0.68) &  -0.01 (0.94)  \\
\hdashline
har & -0.06 (0.39) & 0.25 (0.16) & -0.15 (0.18) &  -0.21 (0.30)  \\
\hdashline
lar & -0.07 (0.20) & 0.06 (0.66) & 0.03 (0.74) & -0.05 (0.76)\\ 
\hdashline
joyemo & -0.02 (0.63) & -0.05 (0.42)  & -0.02 (0.59) & 0.04 (0.66)   \\
\hdashline
scsemo & -0.01 (0.97) &  0.73 (0.49) & -0.50 (0.17) &  -0.12 (0.91)  \\
\hdashline
meeting & -0.01 (0.96) & -0.01 (0.97)  & 0.15 (0.29) &  0.04 (0.82)  \\
\hdashline
failure event & 0.07 (0.18) & 0.04 (0.69)  & 0.05 (0.63) & 0.14 (0.25)   \\
\hline
stress & 0.14 (\textbf{0.01}) &  0.27 (\textbf{<0.001}) & -0.05 (0.74) &  0.01 (0.93)  \\
\hdashline
sleep & -0.04 (0.48)  &  -0.10 (0.32) & -0.03 (0.75) &  0.10 (0.53)  \\
\hdashline
hurry & -0.08 (0.25) &  0.20 (0.42) & -0.30 (0.06) &  0.01 (0.96)  \\
\hdashline
interruptions & 0.01 (0.97) &  -0.32 (0.29) & 0.26 (0.06) &  0.01 (0.92)  \\
\hdashline
ineffective & -0.04 (0.58) & 0.59 (0.17)  & -0.11 (0.39) &  -0.04 (0.79)  \\
\hdashline
independence & -0.02 (0.81) &  0.02 (0.95) & 0.18 (0.31) & 0.02 (0.99)   \\
\hline
Marginal $R^2$ & 0.03 &  0.19 & 0.20 &  0.05  \\
Conditional $R^2$ & 0.52 &  0.21 & 0.20 &  0.05  \\
\hline
Null Model C. $R^2$ & 0.49 & 0  & 0 & 0 &   \\
\end{tabular}
\end{table}


\subsection{RQ4 - \RQFour}\label{sec:interviews}






\paragraph{Motivation}
We wanted to better understand the reason behind the numbers gathered with the daily questionnaire. With interviews, we also hoped to understand better what happened when a particular event occured. For example, whether meetings with the customer were attended by all developers, and what actions a developer had to do when tests for production failed. We also wanted to explore how instant messaging was used in the project, to possibly offer some explanations of our results. Finally, we asked questions related to emoticon and emoji usage, to better understand their usage and meaning in the project chat logs.

\paragraph{Experience Sampling procedure}
All three interviewees, university graduates themselves, mentioned the primary motivation for answering the questionnaire was to offer helpful contributions to science. When asked of the possibility for minor rewards, such as movie tickets, developer 1 said: ``I don't believe that those movie ticket thingies motivate working people. It is not about monetary compensation''. Email messages were also described to be a good way of sharing the link to the questionnaire by all interviewees, as having one's email client open at work was described as being part of the job.

\paragraph{Leadership style, company culture, way of working with respect to the questionnaire}
The project manager described their leadership style to be more facilitating and supportive. In practice, it meant that nobody was ever assigned to specific tasks, but that the developers chose their tasks from a list for the next sprint. The project manager expanded on this: ``Probably a manager wouldn't fare long at the company, who would be saying you do this, you do this and so on''. Both developers expressed that the project team had plenty of independence for making decisions and that the employer did not intervene in day to day decisions.

Furthermore, the project manager told us that the guidelines for developers were to commit small logical changes. Both of the developers backed this up in their interviews, and perhaps as a result, neither of the developers recalled any bigger merge conflicts. We believe this is an important context worth mentioning for the models in the previous subsections. 

\paragraph{Hurry}
Overall, both developers described the project as being without much time pressure. Developer 2 explained: ``I would say that at a general level there was never a terrible hurry... I never felt like somebody was looking over my shoulder; that exactly this task should be ready by a deadline. I knew that if it wouldn't be completed, nothing too terrible would happen.''. Furthermore, both developers described that, while hard deadlines did exist, the needed features were always ready well before this deadline. In the words of developer 1: ``I never felt when we were going to production, that the project is going to cause so much hurry, but well, the version we had is already good enough.''.

Developer 1 offered this after-the-fact explanation: ``Well, I believe, in this project, the feeling of hurry, has been precisely that you don't have time to develop, but 70\% of your time is going to everything else. When you are not in a hurry, you have seven and a half hours to code''. They also further explained ``When I feel like I have to get something done... I don't partake in internal educational events or other training, but I focus on developing the project. And otherwise, maybe I focus more on developing features rather than general project work''. Developer 1 also mentioned that writing tests is a part that could be easily skipped when feeling hurried: ``... The feeling of hurry starts to come when I am implementing tests. But you still have to write the tests.''. The developer thus wrote the tests, but it felt like a part that could be skipped in a pinch.

\paragraph{Role of instant messaging}
All three interviewees agreed that project chat was used for communicating work and technical aspects the overwhelming majority of the time. Another company-wide chat mechanism exists for discussions related to free time, which was not part of our data sources. Two of the interviewees expanded that employees were urged to discuss technical aspects of work specifically on chat over and alongside face-to-face discussions. The benefits mentioned were traces to communications, coordination of expertise with everyone having the same access to information, better focus without interruptions in the shared working space (as opposed to face-to-face communication), and that the team would be aware of issues and solutions related to current events. One example topic for discussion would be for example, whether to integrate a specific new test automation tool to be part of the development process.

One negative consequence of the chat system was mentioned by a single respondent. Private messages from the chat system were seen as interrupting, as they felt there was a higher urgency to respond since a response was demanded specifically from them. This would be the case when the respondent was seen as an expert on some topic, and their opinion and expertise was valued and demanded by the person sending the private message. We want to note that our quantitative data does not include private messages.

According to the developers, some of the emoji used were quite specific to the context and were related to the humor in the project. For example, emojis related to parrots (e.g., ``partyparrot'') were used when things went well or the developer felt something was accomplished. Emojis related to shoveling and a car jack were used when problems arose. The supporting element of the instant messaging channel in relation to the usage of emoji was highlighted by developer 1: ``in those moments when you felt frustrated or irritated, then you would seek support with ``in the trenches''-kind of humor''. We also note that we used this information on the classification of emoji for the quantitative analysis described above.

\paragraph{Job Events}
The project manager described the meetings with the customer in this project as ``very long'', with meetings usually taking three hours. The meetings were ``open meetings''. The project manager further explained that their goal was to circulate the developers to the meetings as they were needed based on their expertise related to the project. For example, when the topic would be a feature, only those who had developed the feature would be in attendance during that part of the meeting. However, the project manager was present from start to finish in the meetings. The project manager further elaborated on this: ``Oh well, the developer does not want to sit in the meetings''.

Neither of the developers could recall situations in which they had to extensively prepare for the meetings. Both developers agreed that some preparation was needed, but it only required thinking about how to demonstrate and what to say about the features they had developed. Developer 2 described the preparation as solely consisting of looking at the agenda, and knowing which development branch in the version control system was the right one for the demonstration when needed. Developer 1 said that the continuous deployment eased the meetings: ``The new code went to the customers' environment, so they could go and use it. I never needed Powerpoint presentations''.

The project manager had the poorest recollections about whether production tests had failed, as significant problems related to hosting the service arose during our study period. However, neither of the developers shared these recollections, perhaps in part because, for hosting and optimization related issues, extra personnel from the operation team outside of the normal development team were involved. While an instant reaction was demanded from the developers, neither of them saw these as particularly bad. Developer 1 explicated: ``It was never a catastrophe, as it only meant that updates to the staging environment would stop and production would not be updated the next morning. Those whose code changes broke the build usually started to fix it as soon as possible. Usually, it was not a big deal.''.

\section{Discussion}\label{sec:discussion}
Ultimately, the main finding of our study is that predicting well-being strongly depends on the individual. While the marginal $R^2$ value did not rise above 0.26 in the models of the individual, such lower $R^2$ values have been reported in more technical studies as well. For example, depending on the project studied, bug prediction models have achieved $R^2$ values in the 0.20's \cite{giger2011comparing, d2010extensive}. 
Is our study a negative result? On the general level, it is as we cannot find shared predictors that would work on all individuals. But on the individual level, it is not as individual predictors were in line with some past work.  

We cannot establish strong links between repository variables and our questionnaire variables related to well-being. We also do not see the links we had between the questionnaire and software repository variables with logistic regression in our prior work with the same dataset \citep{kuutila2018using}. We think this is mostly because of the additional control variables we used as random effects in our model. The main random effect explaining the majority of the variation is the respondent ID in the generalized linear mixed effects models. Our general models for prediction shown in Sections \ref{sec:rsc2} and \ref{sec:rsc3} would look much closer to our previous work if we had not controlled for the individual.

Additionally, repository data are inherently incomplete. However, in some ways repository data will always be incomplete, as \cite{aranda2009secret} have noted: ``the histories of even simple bugs are strongly dependent on social, organizational, and technical knowledge that cannot be solely extracted through automation of electronic repositories, and that such automation provides incomplete and often erroneous accounts of coordination.''. Therefore, repositories always reflect only part of software engineering work actions. Furthermore, events outside work will influence how people feel and sleep, which can influence the questionnaire answers.

Our results are in line with some previous negative results on sentiment analysis studies, e.g. \cite{jongeling2017negative, lin2018sentiment}. Even under laboratory conditions, valence explained 27\%, and arousal 0.5\% of perceived progress in the software development task \citep{girardi2020recognizing}, which is comparable to our productivity measure and the models made with individual data in Table \ref{tab:rq3}. However, we did not find a link between positive valence measured from the chat system and our measured productivity.

In general, the interviews demonstrated that no big deadline pressure or prolonged time pressures were felt during the project, though variance among the answers during the project can be seen in Figs. 1 and 2 in our prior work \citep{kuutila2018daily}. Observing distress and time pressure could be easier when they are more frequent in the software project. Software projects having less time pressure using agile methods are also in line with results from our prior literature review \citep{kuutila2020time}.


We also observe that sending more messages to instant messaging chat was not tied to any clear negative effects. This finding is contrary to some previous work in the information technology field \citep{cameron2005unintended, sykes2011interruptions} where instant messaging was linked to more negative outcomes. The link to more interruptions reported by \cite{sykes2011interruptions} was also reported by one developer during our interviews, but only when using private messages rather than the project-wide chat. Based on the evidence gathered in this study, we believe that using instant messaging applications during software development projects can be beneficial if it is used as a collaborative tool to coordinate expertise, rather than for delivering commands or checking up on whether someone is working. A more facilitating leadership style and a company culture that allowed more independent decisions seemed to be a key contextual difference in this project compared to prior studies.


While the sentiment analysis we performed is quite rudimentary, we demonstrated some links between well-being and variables related to sentiment and emoticon usage. In Table~\ref{tab:t1} positive valence has a positive coefficient with independence. Moreover, one novel aspect of our work is the usage of emoticons and emoji, in Table~\ref{tab:t1} emoticons and emoji related to sadness, confusion, and surprise were statistically significant predictors with regards to independence and hurry.

Finally, we think that one point raised in the interviews is interesting and could be considered in future experience sampling studies. One of the developers mentioned feeling hurried especially when they did not have time for programming and had to do tasks other than development. Such tasks could be related to design, job training and quality assurance. Depending on the project context, one question in a future questionnaire could ask how the developer divided their time between different tasks.


\section{Threats to validity}\label{sec:threats}
\subsection{Internal validity}\label{sec:internal}

The interviews were conducted a considerable amount of time after the questionnaire, and partly because of this we could not interview all the developers who answered the questionnaire. However, the ones interviewed are also some of the ones with the highest response rates to the questionnaire. We tried to help remembrance by sending individual level graphs of the questionnaire answers to the interviewees. We also quantified the interviewees answers, to see how uniform the answers to questions were. Time of the week and month when answering also can influence the answer, which we tried to control with variables in the generalized mixed effects linear models. Other individual traits such as seniority and gender can have an effect, but due to anonymity issues and a low sample size, we do not report these. Experiences and events not related to work can also influence well-being. Thus confounding variables can have an effect on our mixed-effects models.

With regards to generalized linear mixed models, \cite{Bolker2020} collected an encompassing discussion on how to decide whether a variable is fixed or random for generalized linear mixed effects models. \cite{crawley2002statistical} advocated using variables as fixed effects when there not enough levels inside random effects, and \cite{Bolker2020} further sees six levels inside a random effect as the absolute minimum. Thus the levels inside random effects (weekday, respondent ID) can have an effect on our models.

The complexity of random effects structures together with sample size influence model convergence \cite{barr2013random}. Indeed we did have some convergence issues specifically when producing models for individuals where the sample size is lower than the general model. In our case, we simplified the random effects structure by using different moving averages for auto-correlation that helped to get rid of some convergence issues.

\subsection{External validity}\label{sec:external}

The questionnaire was only administered at a single software company with a single software project. This diminishes the generalizability of our results. We tried to contextualize our study partly with the interviews performed in section \ref{sec:interviews}. Major context factors include the company culture, which was described as facilitating and allowing independence for developers, and moreover, without major time pressures. Other contextual factors include an agile way of working, pushing code to production daily, as well as having no big interrogations. We believe our results would be replicable in such a context. However, our study is just one project in a single company, in a single country, and hence, how these different contexts alter the results is yet to be discovered.

\subsection{Construct validity}\label{sec:construct}

The sentiment analysis we performed is rudimentary, mainly because the development team used the Finnish language for instant messages. This severely limited the choice of sentiment analysis tools we could use for this study. The valence lexicon used is not widely known. However, we decided to use it because it is developed specifically for the software engineering context. Studying company-specific jargon would improve the validity of the constructs produced by sentiment analysis, but doing it on a large scale would be a study on its own. We did take some information about the emoticons used into account acquired in the interviews.

Debate on the usage of single-item measures in experience sampling studies exists. Specifically, \cite{rossiter2002c} argues for the validity of ``doubly concrete'' constructs in single-item measurements, that is constructs for which the object and attribute of measurement are unambiguous and clear for the raters. Evidence supporting this view is also presented by multitude of other studies, e.g. \cite{bergkvist2009tailor} and  \cite{wanous1997overall}. More discussion on the subject, including both supporting and contradictory evidence, can be found in an article by \cite{fisher2012using}. Based on the evidence, \cite{fisher2012using} see single-item measurements more valid when they are ``straight forward unidimensional constructs in terms of current or very recent experience'', rather than complicated constructs that are rated retrospectively over a longer time span.



\section{Conclusions}\label{sec:conclusions}
To our knowledge, we present a highly novel study. We observe software developers' well-being with experience sampling over a period of eight months. Additionally, we explore the relationship between well-being and metrics mined from software repositories. If a strong link between well-being and software repositories could be established, this would mean that automated well-being monitoring of software developers would be possible.  

Our results show that developers' well-being varied individually rather than in a collective manner.  We found that software engineering actions (fixed effects) mined mainly from software repositories are not good general predictors of well-being or productivity. Rather it is the individual (modeled as a random effect) that explains differences in well-being and productivity. We further investigated the individuals and found that models of well-being and productivity developed per individual performed better than general models. For example, the top general model had a marginal $R^2$ value of 0.02 while in the individual models top marginal $R^2$ value was 0.26. Thus, adage about predicting ``some of the people some of the time'' holds \citep{bem1974predicting}. 

Future studies on this topic should be improved. A higher number of respondents should be used. However, convincing larger groups to respond to daily surveys over periods of several months is likely to be challenging. Perhaps, the time duration for the survey responses could be shorter, e.g. a month, if the number of individuals responding could be increased to tens of developers. With the increased number of individuals, one could meaningfully study if the individual differences in well-being and productivity that we observed are due to different roles, e.g. senior versus junior developers could have different well-being predictors in software repositories. If one could collect responses from hundreds of developers, then perhaps even personality types could be taken into account\citep{eysenck2020junior}. 

Future studies in software engineering using experience sampling also offer interesting possibilities. Experience sampling can be used to study a multitude of factors related to software engineering. These include the effects of different kinds of processes, techniques, and ways related to software development work, such as the adoption of agile, teleworking, resistance to change, and organizational justice. We also believe that replicating well-being studies in different software development contexts is beneficial, to better understanding contextual factors.



\begin{acknowledgements}
The first, second and third author have been supported by Academy
of Finland grant 298020. The first author has been supported by
Kaute-foundation.
\end{acknowledgements}

\bibliographystyle{spbasic}      
\bibliography{bibliography}   

\begin{thebibliography}{107}
\providecommand{\natexlab}[1]{#1}
\providecommand{\url}[1]{{#1}}
\providecommand{\urlprefix}{URL }
\expandafter\ifx\csname urlstyle\endcsname\relax
  \providecommand{\doi}[1]{DOI~\discretionary{}{}{}#1}\else
  \providecommand{\doi}{DOI~\discretionary{}{}{}\begingroup
  \urlstyle{rm}\Url}\fi
\providecommand{\eprint}[2][]{\url{#2}}

\bibitem[{Adams(2015)}]{adams2015conducting}
Adams WC (2015) Conducting semi-structured interviews. Handbook of Practical
  Program Evaluation pp 492--505

\bibitem[{{\AA}kerstedt et~al.(2002){\AA}kerstedt, Knutsson, Westerholm,
  Theorell, Alfredsson, and Kecklund}]{aakerstedt2002sleep}
{\AA}kerstedt T, Knutsson A, Westerholm P, Theorell T, Alfredsson L, Kecklund G
  (2002) Sleep disturbances, work stress and work hours: a cross-sectional
  study. Journal of psychosomatic research 53(3):741--748

\bibitem[{Alliger and Williams(1993)}]{alliger1993using}
Alliger GM, Williams KJ (1993) Using signal-contingent experience sampling
  methodology to study work in the field: A discussion and illustration
  examining task perceptions and mood. Personnel Psychology 46(3):525--549

\bibitem[{Anderson and Gerbing(1984)}]{anderson1984effect}
Anderson JC, Gerbing DW (1984) The effect of sampling error on convergence,
  improper solutions, and goodness-of-fit indices for maximum likelihood
  confirmatory factor analysis. Psychometrika 49(2):155--173

\bibitem[{Aranda and Venolia(2009)}]{aranda2009secret}
Aranda J, Venolia G (2009) The secret life of bugs: Going past the errors and
  omissions in software repositories. In: 2009 IEEE 31st International
  Conference on Software Engineering, IEEE, pp 298--308

\bibitem[{Bakker and Demerouti(2007)}]{bakker2007job}
Bakker AB, Demerouti E (2007) The job demands-resources model: State of the
  art. Journal of managerial psychology 22(3):309--328

\bibitem[{Bakker et~al.(2005)Bakker, Demerouti, and Euwema}]{bakker2005job}
Bakker AB, Demerouti E, Euwema MC (2005) Job resources buffer the impact of job
  demands on burnout. Journal of occupational health psychology 10(2):170

\bibitem[{Barr et~al.(2013)Barr, Levy, Scheepers, and Tily}]{barr2013random}
Barr DJ, Levy R, Scheepers C, Tily HJ (2013) Random effects structure for
  confirmatory hypothesis testing: Keep it maximal. Journal of memory and
  language 68(3):255--278

\bibitem[{Barton(2009)}]{barton2009mumin}
Barton K (2009) Mumin: multi-model inference. http://r-forge r-project
  org/projects/mumin/

\bibitem[{Bem and Allen(1974)}]{bem1974predicting}
Bem DJ, Allen A (1974) On predicting some of the people some of the time: The
  search for cross-situational consistencies in behavior. Psychological review
  81(6):506

\bibitem[{Bergkvist and Rossiter(2009)}]{bergkvist2009tailor}
Bergkvist L, Rossiter JR (2009) Tailor-made single-item measures of doubly
  concrete constructs. International Journal of Advertising 28(4):607--621

\bibitem[{Bjarnason et~al.(2014)Bjarnason, Hess, Svensson, Regnell, and
  Doerr}]{bjarnason2014reflecting}
Bjarnason E, Hess A, Svensson RB, Regnell B, Doerr J (2014) Reflecting on
  evidence-based timelines. IEEE software 31(4):37--43

\bibitem[{Boehm et~al.(1981)}]{boehm1981software}
Boehm BW, et~al. (1981) Software engineering economics, vol 197. Prentice-hall
  Englewood Cliffs (NJ)

\bibitem[{Bolker et~al.(2020)}]{Bolker2020}
Bolker B, et~al. (2020) Glmm faq.
  \urlprefix\url{https://bbolker.github.io/mixedmodels-misc/glmmFAQ.html#should-i-treat-factor-xxx-as-fixed-or-random}

\bibitem[{Bollen(2001)}]{BOLLEN20017282}
Bollen K (2001) Indicator: Methodology. In: Smelser NJ, Baltes PB (eds)
  International Encyclopedia of the Social \& Behavioral Sciences, Pergamon,
  Oxford, pp 7282 -- 7287,
  \doi{https://doi.org/10.1016/B0-08-043076-7/00709-9},
  \urlprefix\url{http://www.sciencedirect.com/science/article/pii/B0080430767007099}

\bibitem[{Bruckhaus et~al.(1996)Bruckhaus, Madhavii, Janssen, and
  Henshaw}]{bruckhaus1996impact}
Bruckhaus T, Madhavii N, Janssen I, Henshaw J (1996) The impact of tools on
  software productivity. IEEE Software 13(5):29--38

\bibitem[{Brumby et~al.(2019)Brumby, Janssen, and
  Mark}]{brumby2019interruptions}
Brumby DP, Janssen CP, Mark G (2019) How do interruptions affect productivity?
  In: Rethinking Productivity in Software Engineering, Springer, pp 85--107

\bibitem[{Cameron and Webster(2005)}]{cameron2005unintended}
Cameron AF, Webster J (2005) Unintended consequences of emerging communication
  technologies: Instant messaging in the workplace. Computers in Human behavior
  21(1):85--103

\bibitem[{Chrousos and Gold(1992)}]{chrousos1992concepts}
Chrousos GP, Gold PW (1992) The concepts of stress and stress system disorders:
  overview of physical and behavioral homeostasis. JAMA: Journal of the
  American Medical Association 267(9):1244--1252

\bibitem[{Claes et~al.(2018{\natexlab{a}})Claes, M{\"a}ntyl{\"a}, and
  Farooq}]{claes2018use}
Claes M, M{\"a}ntyl{\"a} M, Farooq U (2018{\natexlab{a}}) On the use of
  emoticons in open source software development. In: Proceedings of the 12th
  ACM/IEEE International Symposium on Empirical Software Engineering and
  Measurement, ACM, p~50

\bibitem[{Claes et~al.(2018{\natexlab{b}})Claes, M{\"a}ntyl{\"a}, Kuutila, and
  Adams}]{claes2018programmers}
Claes M, M{\"a}ntyl{\"a} M, Kuutila M, Adams B (2018{\natexlab{b}}) Do
  programmers work at night or during the weekend? In: 2018 IEEE/ACM 40th
  International Conference on Software Engineering (ICSE), IEEE, pp 705--715

\bibitem[{Crawley(2002)}]{crawley2002statistical}
Crawley MJ (2002) Statistical computingan introduction to data analysis using
  S-Plus. 001.6424 C73

\bibitem[{D'Ambros et~al.(2010)D'Ambros, Lanza, and Robbes}]{d2010extensive}
D'Ambros M, Lanza M, Robbes R (2010) An extensive comparison of bug prediction
  approaches. In: 2010 7th IEEE Working Conference on Mining Software
  Repositories (MSR 2010), IEEE, pp 31--41

\bibitem[{De~Kloet et~al.(2005)De~Kloet, Jo{\"e}ls, and
  Holsboer}]{de2005stress}
De~Kloet ER, Jo{\"e}ls M, Holsboer F (2005) Stress and the brain: from
  adaptation to disease. Nature reviews neuroscience 6(6):463

\bibitem[{Demerouti et~al.(2001)Demerouti, Bakker, Nachreiner, and
  Schaufeli}]{demerouti2001job}
Demerouti E, Bakker AB, Nachreiner F, Schaufeli WB (2001) The job
  demands-resources model of burnout. Journal of Applied psychology 86(3):499

\bibitem[{Diaz and Sligo(1997)}]{diaz1997software}
Diaz M, Sligo J (1997) How software process improvement helped motorola. IEEE
  software 14(5):75--81

\bibitem[{Dickersin(1990)}]{dickersin1990existence}
Dickersin K (1990) The existence of publication bias and risk factors for its
  occurrence. Jama 263(10):1385--1389

\bibitem[{Diener et~al.(1999)Diener, Suh, Lucas, and
  Smith}]{diener1999subjective}
Diener E, Suh EM, Lucas RE, Smith HL (1999) Subjective well-being: Three
  decades of progress. Psychological bulletin 125(2):276

\bibitem[{Dirnagl and Lauritzen(2010)}]{dirnagl2010fighting}
Dirnagl U, Lauritzen M (2010) Fighting publication bias: introducing the
  negative results section

\bibitem[{Donald et~al.(2005)Donald, Taylor, Johnson, Cooper, Cartwright, and
  Robertson}]{donald2005work}
Donald I, Taylor P, Johnson S, Cooper C, Cartwright S, Robertson S (2005) Work
  environments, stress, and productivity: An examination using asset.
  International Journal of Stress Management 12(4):409

\bibitem[{Due{\~n}as et~al.(2018)Due{\~n}as, Cosentino, Robles, and
  Gonzalez-Barahona}]{duenas2018perceval}
Due{\~n}as S, Cosentino V, Robles G, Gonzalez-Barahona JM (2018) Perceval:
  Software project data at your will. In: Proceedings of the 40th International
  Conference on Software Engineering: Companion Proceeedings, ACM, pp 1--4

\bibitem[{Elo et~al.(2003)Elo, Lepp{\"a}nen, and Jahkola}]{elo2003validity}
Elo AL, Lepp{\"a}nen A, Jahkola A (2003) Validity of a single-item measure of
  stress symptoms. Scandinavian journal of work, environment \& health pp
  444--451

\bibitem[{Elovainio et~al.(2015)Elovainio, Heponiemi, Jokela, Hakulinen,
  Presseau, Aalto, and Kivim{\"a}ki}]{elovainio2015stressful}
Elovainio M, Heponiemi T, Jokela M, Hakulinen C, Presseau J, Aalto AM,
  Kivim{\"a}ki M (2015) Stressful work environment and wellbeing: What comes
  first? Journal of occupational health psychology 20(3):289

\bibitem[{Eysenck et~al.(2020)Eysenck, Barrett, and
  Saklofske}]{eysenck2020junior}
Eysenck SB, Barrett PT, Saklofske DH (2020) The junior eysenck personality
  questionnaire. Personality and Individual Differences p 109974

\bibitem[{Fanelli(2012)}]{fanelli2012negative}
Fanelli D (2012) Negative results are disappearing from most disciplines and
  countries. Scientometrics 90(3):891--904

\bibitem[{Fisher and To(2012)}]{fisher2012using}
Fisher CD, To ML (2012) Using experience sampling methodology in organizational
  behavior. Journal of Organizational Behavior 33(7):865--877

\bibitem[{Fucci et~al.(2018)Fucci, Scanniello, Romano, and
  Juristo}]{fucci2018need}
Fucci D, Scanniello G, Romano S, Juristo N (2018) Need for sleep: the impact of
  a night of sleep deprivation on novice developers' performance. IEEE
  Transactions on Software Engineering

\bibitem[{Giger et~al.(2011)Giger, Pinzger, and Gall}]{giger2011comparing}
Giger E, Pinzger M, Gall HC (2011) Comparing fine-grained source code changes
  and code churn for bug prediction. In: Proceedings of the 8th Working
  Conference on Mining Software Repositories, pp 83--92

\bibitem[{Girardi et~al.(2020)Girardi, Novielli, Fucci, and
  Lanubile}]{girardi2020recognizing}
Girardi D, Novielli N, Fucci D, Lanubile F (2020) Recognizing developers'
  emotions while programming. In: Proceedings of the ACM/IEEE 42nd
  International Conference on Software Engineering, pp 666--677

\bibitem[{Goldberg and Blackwell(1970)}]{goldberg1970psychiatric}
Goldberg DP, Blackwell B (1970) Psychiatric illness in general practice: a
  detailed study using a new method of case identification. Br med J
  2(5707):439--443

\bibitem[{Graziotin et~al.(2015)Graziotin, Wang, and
  Abrahamsson}]{graziotin2015understanding}
Graziotin D, Wang X, Abrahamsson P (2015) Understanding the affect of
  developers: theoretical background and guidelines for psychoempirical
  software engineering. In: Proceedings of the 7th International Workshop on
  Social Software Engineering, ACM, pp 25--32

\bibitem[{Harris(1989)}]{harris1989nurse}
Harris PE (1989) The nurse stress index. Work \& Stress 3(4):335--346

\bibitem[{Hassan(2009)}]{hassan2009predicting}
Hassan AE (2009) Predicting faults using the complexity of code changes. In:
  Proceedings of the 31st International Conference on Software Engineering,
  IEEE Computer Society, pp 78--88

\bibitem[{Hove and Anda(2005)}]{hove2005experiences}
Hove SE, Anda B (2005) Experiences from conducting semi-structured interviews
  in empirical software engineering research. In: 11th IEEE International
  Software Metrics Symposium (METRICS'05), IEEE, pp 10--pp

\bibitem[{Hyndman et~al.(2007)Hyndman, Khandakar et~al.}]{hyndman2007automatic}
Hyndman RJ, Khandakar Y, et~al. (2007) Automatic time series for forecasting:
  the forecast package for R. 6/07, Monash University, Department of
  Econometrics and Business Statistics

\bibitem[{Ilies and Judge(2004)}]{ilies2004experience}
Ilies R, Judge TA (2004) An experience-sampling measure of job satisfaction and
  its relationships with affectivity, mood at work, job beliefs, and general
  job satisfaction. European journal of work and organizational psychology
  13(3):367--389

\bibitem[{Jenkins et~al.(1988)Jenkins, Stanton, Niemcryk, and
  Rose}]{jenkins1988scale}
Jenkins CD, Stanton BA, Niemcryk SJ, Rose RM (1988) A scale for the estimation
  of sleep problems in clinical research. Journal of clinical epidemiology
  41(4):313--321

\bibitem[{Jongeling et~al.(2015)Jongeling, Datta, and
  Serebrenik}]{jongeling2015choosing}
Jongeling R, Datta S, Serebrenik A (2015) Choosing your weapons: On sentiment
  analysis tools for software engineering research. In: 2015 IEEE International
  Conference on Software Maintenance and Evolution (ICSME), IEEE, pp 531--535

\bibitem[{Jongeling et~al.(2017)Jongeling, Sarkar, Datta, and
  Serebrenik}]{jongeling2017negative}
Jongeling R, Sarkar P, Datta S, Serebrenik A (2017) On negative results when
  using sentiment analysis tools for software engineering research. Empirical
  Software Engineering 22(5):2543--2584

\bibitem[{Karasek(1990)}]{karasek1990healthy}
Karasek R (1990) Healthy work. Stress, productivity, and the reconstruction of
  working life

\bibitem[{Karasek et~al.(1998)Karasek, Brisson, Kawakami, Houtman, Bongers, and
  Amick}]{karasek1998job}
Karasek R, Brisson C, Kawakami N, Houtman I, Bongers P, Amick B (1998) The job
  content questionnaire (jcq): an instrument for internationally comparative
  assessments of psychosocial job characteristics. Journal of occupational
  health psychology 3(4):322

\bibitem[{Kimhy et~al.(2006)Kimhy, Delespaul, Corcoran, Ahn, Yale, and
  Malaspina}]{kimhy2006computerized}
Kimhy D, Delespaul P, Corcoran C, Ahn H, Yale S, Malaspina D (2006)
  Computerized experience sampling method (esmc): assessing feasibility and
  validity among individuals with schizophrenia. Journal of psychiatric
  research 40(3):221--230

\bibitem[{Krippendorff(1980)}]{krippendorff1980reliability}
Krippendorff K (1980) Reliability. Wiley Online Library

\bibitem[{Krippendorff(2011)}]{krippendorff2011computing}
Krippendorff K (2011) Computing krippendorff's alpha-reliability.
  \urlprefix\url{https://repository.upenn.edu/asc_papers/43/}

\bibitem[{Kross et~al.(2013)Kross, Verduyn, Demiralp, Park, Lee, Lin, Shablack,
  Jonides, and Ybarra}]{kross2013facebook}
Kross E, Verduyn P, Demiralp E, Park J, Lee DS, Lin N, Shablack H, Jonides J,
  Ybarra O (2013) Facebook use predicts declines in subjective well-being in
  young adults. PloS one 8(8):e69841

\bibitem[{Kuutila et~al.(2018{\natexlab{a}})Kuutila, M{\"a}ntyl{\"a}, Claes,
  and Elovainio}]{kuutila2018daily}
Kuutila M, M{\"a}ntyl{\"a} MV, Claes M, Elovainio M (2018{\natexlab{a}}) Daily
  questionnaire to assess self-reported well-being during a software
  development project. In: Proceedings of the 3rd International Workshop on
  Emotion Awareness in Software Engineering, ACM, pp 39--43

\bibitem[{Kuutila et~al.(2018{\natexlab{b}})Kuutila, M{\"a}ntyl{\"a}, Claes,
  Elovainio, and Adams}]{kuutila2018using}
Kuutila M, M{\"a}ntyl{\"a} MV, Claes M, Elovainio M, Adams B
  (2018{\natexlab{b}}) Using experience sampling to link software repositories
  with emotions and work well-being. In: Proceedings of the 12th ACM/IEEE
  International Symposium on Empirical Software Engineering and Measurement,
  ACM, p~29

\bibitem[{Kuutila et~al.(2020{\natexlab{a}})Kuutila, M{\"a}ntyl{\"a}, Farooq,
  and Ma{\"e}lick}]{kuutila2020time}
Kuutila M, M{\"a}ntyl{\"a} M, Farooq U, Ma{\"e}lick C (2020{\natexlab{a}}) Time
  pressure in software engineering: A systematic review. Information and
  Software Technology 121:106257,
  \doi{https://doi.org/10.1016/j.infsof.2020.106257},
  \urlprefix\url{http://www.sciencedirect.com/science/article/pii/S0950584920300045}

\bibitem[{Kuutila et~al.(2020{\natexlab{b}})Kuutila, M{\~a}ntyl{\~a}, and
  Claes}]{kuutila2020chat}
Kuutila M, M{\~a}ntyl{\~a} MV, Claes M (2020{\natexlab{b}}) Chat activity is a
  better predictor than chat sentiment on software developers productivity. In:
  Proceedings of the IEEE/ACM 42nd International Conference on Software
  Engineering Workshops, pp 553--556

\bibitem[{Lenberg et~al.(2015)Lenberg, Feldt, and
  Wallgren}]{lenberg2015behavioral}
Lenberg P, Feldt R, Wallgren LG (2015) Behavioral software engineering: A
  definition and systematic literature review. Journal of Systems and software
  107:15--37

\bibitem[{Lin et~al.(2018)Lin, Zampetti, Bavota, Di~Penta, Lanza, and
  Oliveto}]{lin2018sentiment}
Lin B, Zampetti F, Bavota G, Di~Penta M, Lanza M, Oliveto R (2018) Sentiment
  analysis for software engineering: How far can we go? In: Proceedings of the
  40th International Conference on Software Engineering, pp 94--104

\bibitem[{Linton(2004)}]{linton2004does}
Linton SJ (2004) Does work stress predict insomnia? a prospective study.
  British Journal of Health Psychology 9(2):127--136

\bibitem[{Liu(2009)}]{liu2009handbook}
Liu B (2009) Handbook chapter: Sentiment analysis and subjectivity. handbook of
  natural language processing. Handbook of Natural Language Processing Marcel
  Dekker, Inc New York, NY, USA

\bibitem[{M{\"a}ntyl{\"a} et~al.(2014)M{\"a}ntyl{\"a}, Petersen, Lehtinen, and
  Lassenius}]{mantyla2014time}
M{\"a}ntyl{\"a} MV, Petersen K, Lehtinen TO, Lassenius C (2014) Time pressure:
  a controlled experiment of test case development and requirements review. In:
  Proceedings of the 36th International Conference on Software Engineering,
  ACM, pp 83--94

\bibitem[{M{\"a}ntyl{\"a} et~al.(2017)M{\"a}ntyl{\"a}, Novielli, Lanubile,
  Claes, and Kuutila}]{mantyla2017bootstrapping}
M{\"a}ntyl{\"a} MV, Novielli N, Lanubile F, Claes M, Kuutila M (2017)
  Bootstrapping a lexicon for emotional arousal in software engineering. In:
  2017 IEEE/ACM 14th International Conference on Mining Software Repositories
  (MSR), IEEE, pp 198--202

\bibitem[{Mark et~al.(2008)Mark, Gudith, and Klocke}]{mark2008cost}
Mark G, Gudith D, Klocke U (2008) The cost of interrupted work: more speed and
  stress. In: Proceedings of the SIGCHI conference on Human Factors in
  Computing Systems, ACM, pp 107--110

\bibitem[{Marsh et~al.(1996)Marsh, Balla, and Hau}]{marsh1996evaluation}
Marsh HW, Balla JR, Hau KT (1996) An evaluation of incremental fit indices: A
  clarification of mathematical and empirical properties. Advanced structural
  equation modeling: Issues and techniques pp 315--353

\bibitem[{McEwen and Sapolsky(1995)}]{mcewen1995stress}
McEwen BS, Sapolsky RM (1995) Stress and cognitive function. Current opinion in
  neurobiology 5(2):205--216

\bibitem[{Miller et~al.(2016)Miller, Thebault-Spieker, Chang, Johnson, Terveen,
  and Hecht}]{miller2016blissfully}
Miller HJ, Thebault-Spieker J, Chang S, Johnson I, Terveen L, Hecht B (2016)
  “blissfully happy” or “ready to fight”: Varying interpretations of
  emoji. In: Tenth International AAAI Conference on Web and Social Media

\bibitem[{Miner et~al.(2005)Miner, Glomb, and Hulin}]{miner2005experience}
Miner A, Glomb T, Hulin C (2005) Experience sampling mood and its correlates at
  work. Journal of Occupational and Organizational Psychology 78(2):171--193

\bibitem[{Mockus et~al.(2002)Mockus, Fielding, and Herbsleb}]{mockus2002two}
Mockus A, Fielding RT, Herbsleb JD (2002) Two case studies of open source
  software development: Apache and mozilla. ACM Transactions on Software
  Engineering and Methodology (TOSEM) 11(3):309--346

\bibitem[{Murphy-Hill et~al.(2019)Murphy-Hill, Jaspan, Sadowski, Shepherd,
  Phillips, Winter, Knight, Smith, and Jorde}]{murphy2019predicts}
Murphy-Hill E, Jaspan C, Sadowski C, Shepherd D, Phillips M, Winter C, Knight
  A, Smith E, Jorde M (2019) What predicts software developers' productivity?
  IEEE Transactions on Software Engineering

\bibitem[{Muthén and Muthén(2017)}]{UnusualTLIvalues}
Muthén LK, Muthén BO (2017) Unusual tli values.
  \urlprefix\url{https://www.statmodel.com/download/TLI.pdf}

\bibitem[{Nagy(2002)}]{nagy2002using}
Nagy MS (2002) Using a single-item approach to measure facet job satisfaction.
  Journal of occupational and organizational psychology 75(1):77--86

\bibitem[{Nakagawa and Schielzeth(2013)}]{nakagawa2013general}
Nakagawa S, Schielzeth H (2013) A general and simple method for obtaining r2
  from generalized linear mixed-effects models. Methods in ecology and
  evolution 4(2):133--142

\bibitem[{Nan and Harter(2009)}]{nan2009impact}
Nan N, Harter DE (2009) Impact of budget and schedule pressure on software
  development cycle time and effort. IEEE Transactions on Software Engineering
  35(5):624--637

\bibitem[{Novielli et~al.(2018)Novielli, Girardi, and
  Lanubile}]{novielli2018benchmark}
Novielli N, Girardi D, Lanubile F (2018) A benchmark study on sentiment
  analysis for software engineering research. In: 2018 IEEE/ACM 15th
  International Conference on Mining Software Repositories (MSR), IEEE, pp
  364--375

\bibitem[{Pinheiro et~al.(2017)Pinheiro, Bates, DebRoy, Sarkar, Heisterkamp,
  Van~Willigen, and Maintainer}]{pinheiro2017package}
Pinheiro J, Bates D, DebRoy S, Sarkar D, Heisterkamp S, Van~Willigen B,
  Maintainer R (2017) Package ‘nlme’. Linear and nonlinear mixed effects
  models, version 3(1)

\bibitem[{Pitk{\"a}nen(2012)}]{Voikko}
Pitk{\"a}nen H (2012) Voikko - free linquistic software and data for finnish.
  \urlprefix\url{https://voikko.puimula.org/}

\bibitem[{Plutchik(1991)}]{plutchik1991emotions}
Plutchik R (1991) The emotions. University Press of America

\bibitem[{Rossiter(2002)}]{rossiter2002c}
Rossiter JR (2002) The c-oar-se procedure for scale development in marketing.
  International journal of research in marketing 19(4):305--335

\bibitem[{Russell(2003)}]{russell2003core}
Russell JA (2003) Core affect and the psychological construction of emotion.
  Psychological review 110(1):145

\bibitem[{Schermelleh-Engel et~al.(2003)Schermelleh-Engel, Moosbrugger,
  M{\"u}ller et~al.}]{schermelleh2003evaluating}
Schermelleh-Engel K, Moosbrugger H, M{\"u}ller H, et~al. (2003) Evaluating the
  fit of structural equation models: Tests of significance and descriptive
  goodness-of-fit measures. Methods of psychological research online
  8(2):23--74

\bibitem[{Schmidt(2004)}]{schmidt2004analysis}
Schmidt C (2004) The analysis of semi-structured interviews. A companion to
  qualitative research pp 253--258

\bibitem[{Schuler(1980)}]{schuler1980definition}
Schuler RS (1980) Definition and conceptualization of stress in organizations.
  Organizational behavior and human performance 25(2):184--215

\bibitem[{Schulte and Vainio(2010)}]{schulte2010well}
Schulte P, Vainio H (2010) Well-being at work--overview and perspective.
  Scandinavian journal of work, environment \& health pp 422--429

\bibitem[{Schwarz and Clore(1983)}]{schwarz1983mood}
Schwarz N, Clore GL (1983) Mood, misattribution, and judgments of well-being:
  informative and directive functions of affective states. Journal of
  personality and social psychology 45(3):513

\bibitem[{Scollon et~al.(2009)Scollon, Prieto, and
  Diener}]{scollon2009experience}
Scollon CN, Prieto CK, Diener E (2009) Experience sampling: promises and
  pitfalls, strength and weaknesses. In: Assessing well-being, Springer, pp
  157--180

\bibitem[{Semmer et~al.(1996)Semmer, Zapf, and Greif}]{semmer1996shared}
Semmer N, Zapf D, Greif S (1996) Shared job strain: a new approach for
  assessing the validity of job stress measurements. Journal of occupational
  and organizational psychology 69(3):293--310

\bibitem[{Silla and Gamero(2014)}]{silla2014shared}
Silla I, Gamero N (2014) Shared time pressure at work and its health-related
  outcomes: Job satisfaction as a mediator. European Journal of Work and
  Organizational Psychology 23(3):405--418

\bibitem[{Singh and Suar(2013)}]{singh2013health}
Singh P, Suar D (2013) Health consequences and buffers of job burnout among
  indian software developers. Psychological Studies 58(1):20--32

\bibitem[{Snir and Zohar(2008)}]{snir2008workaholism}
Snir R, Zohar D (2008) Workaholism as discretionary time investment at work: An
  experience-sampling study. Applied Psychology 57(1):109--127

\bibitem[{Sonnentag et~al.(1994)Sonnentag, Brodbeck, Heinbokel, and
  Stolte}]{sonnentag1994stressor}
Sonnentag S, Brodbeck FC, Heinbokel T, Stolte W (1994) Stressor-burnout
  relationship in software development teams. Journal of occupational and
  organizational psychology 67(4):327--341

\bibitem[{Sun(2005)}]{sun2005assessing}
Sun J (2005) Assessing goodness of fit in confirmatory factor analysis.
  Measurement and evaluation in counseling and development 37(4):240--256

\bibitem[{Svenson(1993)}]{svenson1993time}
Svenson O (1993) Time pressure and stress in human judgment and decision
  making. Springer Science \& Business Media

\bibitem[{Sykes(2011)}]{sykes2011interruptions}
Sykes ER (2011) Interruptions in the workplace: A case study to reduce their
  effects. International Journal of Information Management 31(4):385--394

\bibitem[{Taipale et~al.(2011)Taipale, Selander, Anttila, and
  N{\"a}tti}]{taipale2011work}
Taipale S, Selander K, Anttila T, N{\"a}tti J (2011) Work engagement in eight
  european countries: The role of job demands, autonomy, and social support.
  International Journal of Sociology and Social Policy 31(7/8):486--504

\bibitem[{Tarafdar et~al.(2007)Tarafdar, Tu, Ragu-Nathan, and
  Ragu-Nathan}]{tarafdar2007impact}
Tarafdar M, Tu Q, Ragu-Nathan BS, Ragu-Nathan T (2007) The impact of
  technostress on role stress and productivity. Journal of Management
  Information Systems 24(1):301--328

\bibitem[{Thompson(2004)}]{thompson2004exploratory}
Thompson B (2004) Exploratory and confirmatory factor analysis. American
  Psychological Association

\bibitem[{Tregubov et~al.(2017)Tregubov, Boehm, Rodchenko, and
  Lane}]{tregubov2017impact}
Tregubov A, Boehm B, Rodchenko N, Lane JA (2017) Impact of task switching and
  work interruptions on software development processes. In: Proceedings of the
  2017 International Conference on Software and System Process, ACM, pp
  134--138

\bibitem[{Tucker and Lewis(1973)}]{tucker1973reliability}
Tucker LR, Lewis C (1973) A reliability coefficient for maximum likelihood
  factor analysis. Psychometrika 38(1):1--10

\bibitem[{Vrijkotte et~al.(2000)Vrijkotte, Van~Doornen, and
  De~Geus}]{vrijkotte2000effects}
Vrijkotte TG, Van~Doornen LJ, De~Geus EJ (2000) Effects of work stress on
  ambulatory blood pressure, heart rate, and heart rate variability.
  Hypertension 35(4):880--886

\bibitem[{Wagner and Ruhe(2018)}]{wagner2018systematic}
Wagner S, Ruhe M (2018) A systematic review of productivity factors in software
  development. arXiv preprint arXiv:180106475

\bibitem[{Wanous et~al.(1997)Wanous, Reichers, and Hudy}]{wanous1997overall}
Wanous JP, Reichers AE, Hudy MJ (1997) Overall job satisfaction: how good are
  single-item measures? Journal of applied Psychology 82(2):247

\bibitem[{West and Hepworth(1991)}]{west1991statistical}
West SG, Hepworth JT (1991) Statistical issues in the study of temporal data:
  Daily experiences. Journal of Personality 59(3):609--662

\bibitem[{Xanthopoulou et~al.(2007)Xanthopoulou, Bakker, Dollard, Demerouti,
  Schaufeli, Taris, and Schreurs}]{xanthopoulou2007job}
Xanthopoulou D, Bakker AB, Dollard MF, Demerouti E, Schaufeli WB, Taris TW,
  Schreurs PJ (2007) When do job demands particularly predict burnout? the
  moderating role of job resources. Journal of managerial psychology
  22(8):766--786

\bibitem[{Xanthopoulou et~al.(2009)Xanthopoulou, Bakker, Demerouti, and
  Schaufeli}]{xanthopoulou2009reciprocal}
Xanthopoulou D, Bakker AB, Demerouti E, Schaufeli WB (2009) Reciprocal
  relationships between job resources, personal resources, and work engagement.
  Journal of Vocational behavior 74(3):235--244

\end{thebibliography}

%
%

\end{document}